\documentclass{rsproca_new}

\usepackage[utf8]{inputenc}
\usepackage[T1]{fontenc}
 \usepackage{amsmath}
 \pdfoutput=1
\usepackage{float}
\usepackage{bm}
\usepackage[normalem]{ulem}
\newcommand{\beginsupplement}{%
        \setcounter{table}{0}
        \renewcommand{\thetable}{S\arabic{table}}%
        \setcounter{figure}{0}
        \renewcommand{\thefigure}{S\arabic{figure}}%
     }

\begin{document}

\title{Estimating the effective reproduction number for heterogeneous models using incidence data}

\author{
D. C. P. Jorge$^{1}$, J. F. Oliveira$^{2}$, J. G. V. Miranda$^{3}$, R. F. S. Andrade$^{2,3}$ and S. T. R. Pinho$^{3}$}

\address{$^{1}$Instituto de Física Teórica, Universidade Estadual Paulista - UNESP, R. Dr. Teobaldo Ferraz 271, São Paulo 01140-070, Brazil\\
$^{2}$Center of Data and Knowledge Integration for Health (CIDACS), Instituto Gonçalo Moniz, Fundação Oswaldo Cruz, Salvador, Bahia, Brazil\\
$^{3}$Instituto de Física, Universidade Federal da Bahia, Salvador, Bahia, Brazil}

\subject{Mathematical modeling, Differential equations, Applied mathematics}


\corres{Daniel Cardoso Pereira Jorge\\
\email{danielcpjorge98@gmail.com}}

\begin{abstract}
The effective reproduction number, $ \mathcal {R} (t) $, is a central point in the study of infectious diseases. It establishes in an explicit way the extent of an epidemic spread process in a population. The current estimation methods for the time evolution of $\mathcal{R}(t)$, using incidence data, rely on the generation interval distribution, $ g (\tau) $, which is usually obtained from empirical data or already known distributions from the literature. However, there are systems, especially highly heterogeneous ones, in which there is a lack of data and an adequate methodology to obtain $g(\tau)$. In this work, we use mathematical models to bridge this gap. We present a general methodology for obtaining an explicit expression of the reproduction numbers and the generation interval distributions provided by an arbitrary compartmental model. Additionally, we present the appropriate expressions to evaluate those reproduction numbers using incidence data. To highlight the relevance of such methodology, we apply it to the spread of Covid-19 in municipalities of the state of Rio de janeiro, Brazil. Using two meta-population models, we estimate the reproduction numbers and the contributions of  each municipality in the generation of cases in all others. Our results point out the importance of mathematical modelling to provide epidemiological meaning of the available data.     
\end{abstract}


\maketitle
\section{Introduction}
The human population gets along with different microorganisms. Some of them lead to transmitted diseases that result in epidemics, even pandemics such as SARS-Cov-2. It is very important to define reliable measures to characterize the spread of those pathogens, both at the beginning and in the course of epidemics.

The reproduction number $\mathcal{R}(t)$ indicates the number of new infections that result from a single infected individual at any time $t$. When it is greater than 1, each infected individual tends to generate more than one infected individuals resulting in the occurrence of an epidemic outbreak. The reproduction number is usually known from its initial value, when all of the population is susceptible, known as the basic reproduction number $\mathcal{R}_0$. Due to the recovery of infected individuals, a portion of the population becomes immune to the disease, thus, herd immunity begins to emerge in the population. This acts as a barrier to the transmission of the disease, so $\mathcal{R}(t)$ takes into account the evolution of the number of susceptible individuals in the population and is usually referred as the effective reproduction number. It provides us with information on how disease transmission occurs during the outbreak it is an important metric to make healthcare managers  aware of the current level of the disease propagation. 

By considering heterogeneities in the homogeneous compartmental models, the reproduction number is affected. Indeed different classes of infected individuals can lead to different contributions for the generation of new infections. A lot of studies have taken the heterogeneity into account when dealing with the basic reproduction number $\mathcal{R}_0$ \cite{Hyman,Brauer}, however, this is not the same for the effective reproduction number, $\mathcal{R}(t)$. 

To study reproduction number, the renewal equation, a Lotka-Euler type equation \cite{Wallinga}, is usually used. This equation emerges from the mathematical models dynamics and can be derived using infection-age models. This type of modelling comes from the study of age-structured population growth and was first used within the scope  of mathematical demography with the aim of modeling the birth dynamics due to the offspring produced by an individual over its lifespan. Analogously, the same methodology can be applied to the infectious diseases scenario by modelling the generation of new infected individuals during the infection period of those previously infected \cite{Martcheva}. Infection-age models are the origins of the widely known SIR and SEIR models developed by Kermack-Mckendrick \cite{Kermack}. Most known compartment ODE models for infectious diseases have their infection-age identical counterparts. The advantage with respect to the infection-age models is that they allow us to have access to the distribution of the infected individuals during the infectious period. 


The estimation of the reproduction number in the course of an outbreak usually rely on the daily count of new cases along with the renewal equation. This approach uses the generation interval distribution, also called the serial interval or generation time distribution, and was popularized by \cite{Wallinga}. Several works indicate how to obtain this distribution through empirical data\cite{Park,Huber,Lessler,Nishiura 2020} or from other known distributions and models \cite{Nishiura,Wallinga,Champredon,Akkouchi,Champredon2}. However, these distributions rely on epidemiological and empirical studies and several systems are very difficult to analyse or even do not have the required data for the envisaged evaluation. Moreover, highly heterogeneous systems do not have a general methodology for estimating the reproduction number and its generation interval distribution \cite{Nishiura 2009,Park 2019, Wesley, Fraser}.


In this study, we use mathematical models to bridge the gap between the reported cases and the reproduction number. Therefore, we present a methodology that derives reproduction numbers and the generation interval distribution for an arbitrary compartment mathematical model. With such methodology, we can investigate highly heterogeneous systems using appropriate models and evaluate their reproduction numbers. With that in mind, apply the method to a meta-population model to  analyse  the  role  of  spatial  heterogeneity  in  the  spreading  of  COVID-19 in  the  Brazilian  territory.


Our work is separated as follow:  in section 2, we introduce heterogeneity by allowing the existence of different groups of individuals in a population that can be sorted into compartments \cite{Hethcote}, considering then a very general heterogeneous model that can be reduced to most models in literature. We use this general formulation to develop our methodology to obtain the reproduction numbers and expressions that correlate it to actual data. Then, in section 3, we  apply it to a specific scenario related to the role of the inter-municipal commuter flow of people and extract its reproduction number and the generation interval distribution. In section 4 and 5 we use actual data of the emerging SARS-Cov-2 coronavirus pandemic in the municipalities of the state of Rio de Janeiro, Brazil, to estimate the reproduction numbers that emerge from the system. Additionally, we reconstruct the time series of cases from the contributions of each municipality in the propagation of the disease. 


\section{Reproduction number in a heterogeneous population}

\subsection{A general infection-age model}

In a heterogeneous population, individuals can be discernible by age-groups, spatial locations, behaviour, different susceptibility to diseases or any other factor that may distinguish them for one another.
In this section, we consider a general heterogeneous model that separates the individuals in $m$ homogeneous compartments. For the purpose of evaluating the reproduction number, we first consider a sub-set of variables encompassing the infected compartments, i.e., those compartments containing individuals that carry the etiologic agent in their organisms. These compartments are denoted by $\bm{x}(t,\tau)=(x_1(t,\tau) , ...,x_n(t,\tau))$, where $t$ and $\tau$ indicate, respectively, the calendar time and the infection-age, the time elapsed since they got the infection. Of course the condition $\bm{x}(t,\tau)=0$ for $\tau >t$ must be satisfied. If $m=m_i+m_0$, where $m_i$ and $m_0$ indicate the number of infected and non-infected compartments in the homogeneous model, $n=m_i$. Therefore, $\bm{x}$ represents the number of individuals in each compartment at a specific calendar time ``$t$'' and at infection age ``$\tau$''. $\bm{x}$ entries can be called the infection-age distributions. The total number of individuals in each compartment, denoted as $X_i(t)$, can be obtained by integrating $x_i(t,\tau)$ with respect of $\tau$ from zero to infinity, where $\bm{X}(t)=(X_1(t) , ...,X_n(t))$. For simplicity, we will be using matrix notation in some equations. The matrices and vectors will be identified by using bold font, whereby $\bm{M}(y,z)= \Big [M_{ij}(y,z) \Big]$. Also, as usual, integrals and derivatives with respect of scalar variables operate componentwise.

Drawing a parallel with Van den Driessche's next generation method \cite{van den Driessche}, we distinguish the new infections from the other compartment flows, defining $\bm{\mathcal{F}}(t)=(\mathcal{F}(t) , ...,\mathcal{F}_n(t))$ as the rate of appearance of new infections and $\bm{\mathcal{V}}(t,\tau)=(\mathcal{V}_1(t,\tau) , ...,\mathcal{V}_n(t,\tau))$ as the rate of transfer into compartments. Therefore, $\bm{\mathcal{F}}(t)$ describes the flow from non-infected compartments into infected ones and depends on $\bm{X}(t)$. On the other hand,$\bm{\mathcal{V}}(t,\tau)$ is related to the flow between infected compartments or from infected to non-infected ones and must depend on $\bm{x}(t,\tau)$. Thus, an usual infection-age model can be written as:

 \begin{align}
\Big( \frac{\partial}{\partial t} +\frac{\partial}{\partial \tau}\Big)\bm{x}(t,\tau)=& -\bm{\mathcal{V}}(t,\tau),\label{Infec_V}\\
\bm{x}(t,\tau=0)=& \bm{\mathcal{F}}(t). \label{Infec_F}
\end{align}

Infection-age models are partial differential equations (PDE) that takes into account the time elapsed since the infection $\tau$. The equations \eqref{Infec_V} and \eqref{Infec_F} can usually be linked to their respective ODE models, and in fact the Kermack-Mckendrick SIR and SEIR models \cite{Kermack} are the special cases of their infection-age correspondents. This modeling allows us to have access to the distribution of the infected during the infectious period, which enable us to estimate the number of active infected individuals and the reproduction number along the outbreak using the reported data. Integrating \eqref{Infec_V} from zero to infinity of  with respect of $\tau$ we obtain:

\begin{equation}\label{correspondece}
    \frac{d}{dt} \bm{X}(t) = \bm{\mathcal{F}}(t) - \int_0 ^{\infty} \bm{\mathcal{V}}(t,\tau) d\tau,
\end{equation}

\noindent which is the correspondent to the sub-set of equations for the infected variables of the ODE model. Therefore, even though we are dealing with an infection-age model, one can simply build $\bm{\mathcal{F}}$ and $\bm{\mathcal{V}}$ from an ODE system and apply the methodology in this work. We demonstrate this for usual ODE models in the Supplementary Material 1.

Since $\mathcal{F}_i(t)$ and $\mathcal{V}_i(t,\tau)$ are well defined, we proceed to solving equation \eqref{Infec_V} using the method of integration along the characteristic line. Thus, the left side of the equation \eqref{Infec_V} indicates that the characteristics of those PDE's are lines of slope 1, which implies that $t=\tau+c$ with $c$ being an arbitrary constant. We fix a point ($t_0$,$\tau_0$) and introduce a variable $\omega$ such that $u_i(s)=x_i(t_0+\omega,\tau_0+\omega)$ are functions that provide the values of the compartment densities along the characteristic. After straightforward calculations \cite{Martcheva}, we obtain 

\begin{equation}\label{solve_V}
    \frac{d}{d\omega} \bm{u}(\omega) = -\bm{\overline{\mathcal{V}}}(\omega),
\end{equation}

\noindent where $\overline{\mathcal{V}}_i(\omega)= \mathcal{V}_i(t_0 +\omega, \tau_0 +\omega)$. In epidemic models, $\Bar{\mathcal{V}}_i(\omega)$ are usually linear equations, making this system easy to solve. In that case, the system can be represented as:

\begin{equation}\label{solve_V_x}
    \frac{d}{d\omega} \bm{u} = -\frac{\partial \; \bm{\overline{\mathcal{V}}}}{\partial \bm{u}} \; \bm{u},
\end{equation}

\noindent where $-\frac{\partial\bm{\overline{\mathcal{V}}}}{\partial \bm{u}}= \Big[ -\frac{\partial}{\partial u_j} \overline{\mathcal{V}}_i(\omega) \Big]$ is the matrix of the linear system. Assuming that there are no infected individuals prior to $t=0$, we only need to take into account the solution where $t>\tau$, leading to $\omega=\tau$, $t=\tau +t_0$ and $\tau_0 = 0$. The solution for a linear system can be written as

\begin{equation}\label{Solution_u}
    \bm{u}(\omega) =  \bm{\overline{\Gamma}}(\omega) \; \bm{u}(0) ,
\end{equation}

\noindent where $\bm{\overline{\Gamma}}(\omega)=\bm{\Gamma}(t,\tau)$, is the fundamental matrix obtained by solving \eqref{solve_V}.  Therefore, we identify $\bm{\mathcal{F}}(t_0)= \bm{u}(0)$ in \eqref{Infec_F}. So, \eqref{Solution_u} becomes

\begin{equation}\label{Solution_V}
    \bm{x}(t,\tau) =  \bm{\Gamma}(t,\tau) \; \bm{\mathcal{F}}(t-\tau).
\end{equation}

 For a linear $\bm{\mathcal{V}}(t,\tau)$, $\bm{\Gamma}(\tau)$ components are exponential functions. Equation \eqref{Solution_V} can also be used to express the solution of a general non-linear $\bm{\mathcal{V}}(t,\tau)$, but if it is not the case, such system will not be considered in this work. On the other hand, we can also assume that $\mathcal{F}_i(t)$ can be written as
 
 \begin{equation}\label{F_het}
    \mathcal{F}_i(t)= \sum_{j}^{n} \int_{0}^{\infty} \Omega_{ij}(t,\tau) \; x_j(t,\tau) d\tau,
\end{equation}

\noindent in which $\bm{\Omega}(t,\tau)$ is related to the generation of infected individuals in "$i$" due to "$j$". If $\bm{\Omega}$ does not depend on $\tau$, as in ODE models, it assumes the form of

\begin{equation}\label{Omega_deriv}
    \bm{\Omega}(t)= \Big[ \frac{\partial}{\partial x_j} \mathcal{F}_i(t) \Big]
\end{equation}

Most system in literature satisfies \eqref{Solution_V} and \eqref{F_het}, thus the method in this work is very general and can be applied to a wide range of models.    


\subsection{Obtaining the reproduction numbers}

To estimate the reproduction number using the reported data of new infections, we need to link them with the equations of the model. This can be done by substituting \eqref{Solution_V} in \eqref{F_het}. Arranging the equation we get

\begin{equation}\label{renewal_het}
    \bm{\mathcal{F}}(t)=\int_{0}^{\infty} \bm{A}(t,\tau) \bm{\mathcal{F}}(t-\tau) d\tau,
\end{equation}

\noindent for 

\begin{equation}\label{A_het}
    \bm{A}(t,\tau)=  \bm{\Omega}(t,\tau) \bm{\Gamma}(t,\tau).
\end{equation}

The functions $A_{ij}(t,\tau)$, analogously to \cite{Nishiura}, are referred as the rate of new infections in $X_i$ due to $X_j$ previous infections at a calendar time $t$ and infection-age $\tau$, whereby $A_{ij}(t,\tau>t)\equiv 0$. So, we can describe the new cases in "$X_i$" from the cases that occurred previously in all the groups. In fact, \eqref{renewal_het} is a general form for the widely known renewal equation, \cite{Nishiura, Wallinga}, actually a sum of renewal equations. Thus, by defining the number of new infections in the "$i$" compartment due to "$j$" previews infections as

\begin{equation}\label{renewal_het_ij}
    \mathcal{J}_{ij}(t)= \int_{0}^{\infty} A_{ij}(t,\tau) \mathcal{F}_j(t-\tau) d\tau,
\end{equation}

\noindent it becomes clear that $\mathcal{F}_i(t)=\sum_j^{n} \mathcal{J}_{ij}(t)$. Thus, it is possible to quantify the influence of the infections occurred in one compartment new ones in another compartment. To obtain the expected number of new infections in $X_i$ a newly infected individual form $X_j$ is expected to generate, thus the reproduction number, we integrate $A_{ij}(t,\tau)$ from zero to infinity with respect to $\tau$, as in \cite{Nishiura}.

\begin{equation}\label{Rt_ij}
    \mathcal{R}_{ij}(t)= \int_0^\infty A_{ij}(t,\tau) d\tau.
\end{equation}

$\bm{\mathcal{R}}$ is usually called the next-generation matrix, where its entries corresponds to the reproduction numbers of the system \cite{Nishiura 2009}. It is remarkable that, for a ODE model where $\bm{\mathcal{V}}$ is linear, the next-generation matrix evaluated at the free-disease fixed point $\bm{\mathcal{R}|}_{x_0}$, is equivalent to the one developed in \cite{van den Driessche}. Naturally from \eqref{Rt_ij}, we define

\begin{equation}\label{g_ij}
    g_{ij}(t,\tau) = \frac{A_{ij}(t,\tau)}{\int_0^\infty A_{ij}(t,\tau) d\tau},
\end{equation}

\noindent where $g_{ij}(t,\tau)$ is normalized and denoted as the generation interval distribution \cite{Nishiura,Wallinga}. Thus, it is related to the flow of individuals between infected compartments and their recover process. We assume a generation interval distribution depending on $t$ and $\tau$, which is general enough to be applied to models whose dynamics can change over time. Therefore, using \eqref{Rt_ij} and \eqref{g_ij} in \eqref{renewal_het_ij} we obtain:

\begin{equation}\label{J_ij_R}
    \mathcal{J}_{ij}(t)=  \mathcal{R}_{ij}(t) \int_0^\infty g_{ij}(t,\tau) \;\mathcal{F}_j(t-\tau) d\tau.
\end{equation}

It is important to highlight that $ \mathcal {R} _ {ij} (t) $ is not necessarily the number of new infected in $ X_i $ generated by $ X_j $. Instead, the meaning of $ \mathcal {R} _ {ij} (t) $ is linked to the generation of new infected ones in ``$i$'', $ \mathcal {F} _i (t) $, due to infected individuals previously generated in ``$ j $'', $ \mathcal {F} _j (t- \tau) $, regardless of what stage of the disease these infected individuals who entered ``$ j $'' are at instant $ t $. That is the case because a new infected individual may generate new infections through out many stages of the disease. In the SEIR model, for example, all of the new infections are in the $E$ compartment, but the individual only becomes infective when it goes to the $I$. To access the reproduction number a newly infected in $X_j$ is expected to generate to all of the other compartments, we can just sum the $\mathcal{R}_{ij}$ over $i$, defining

\begin{equation}\label{R_bar_j}
    \overline{\mathcal{R}}_j(t) = \sum_i ^n\mathcal{R}_{ij}(t).
\end{equation}

\noindent Through out this work, the over-line represents the merge of the first index , in this case, in the form of a sum. This lead us to analogously define $\overline{A}_j= \sum_i ^n A_{ij}$, whereby its integral from zero to infinity with respect to $\tau$ is $\overline{\mathcal{R}}_j$. Thus, we are able to write

\begin{equation}\label{J_bar}
        \overline{\mathcal{J}}_{j}(t)=  \overline{\mathcal{R}}_j(t) \int_0^\infty \overline{g}_{j}(t,\tau) \;\mathcal{F}_j(t-\tau) d\tau,
\end{equation}

\noindent where $\overline{\mathcal{J}}_{j}(t) = \sum_i ^{n} \mathcal{J}_{ij}(t)$ and

\begin{equation}\label{g_bar}
    \overline{g}_{j}(t,\tau) =  \frac{\overline{A}_{j}(t,\tau)}{\int_0^\infty \overline{A}_{j}(t,\tau) d\tau}= \frac{\sum_i ^n \mathcal{R}_{ij} (t) g_{ij}(t,\tau)}{\sum_i ^n \mathcal{R}_{ij} (t) }
\end{equation}

Therefore, $\overline{\mathcal{J}}_j(t)$ is the total number of new infections generated by previews infections in $X_j$. It is interesting that the generation interval distribution $\overline{g}_j(t,\tau)$ takes the form of a 
weighted average in which the $g_{ij}(t,\tau)$ are the weights.

Thus, the implementation of the proposed method in this work amounts to: identifying the terms $\bm{\mathcal{F}}$ and $\bm{\mathcal{V}}$ from a model; using them to find $\bm{\Omega}$ and $\bm{\Gamma}$; obtaining $\bm{A}$ and integrating to get $\bm{\mathcal{R}}$ and $\bm{g}$. Further in this work we present applications of the method and estimations using actual data. Examples of the method in different types of models can be found on the Supplementary Material 1.

\subsection{The total reproduction number}

After obtaining the reproduction numbers of the constituents of the system, we now intend to define a reproduction number for the whole system. The number of new infections in a group, $\mathcal{F}_i(t)$, can be described as a fraction of the total number of cases from all groups, $\mathcal{F}^T(t)=\sum_i ^n \mathcal{F}_i(t)$ such that 

\begin{equation}\label{F_prop}
   \mathcal{F}_i(t)= \alpha_i(t) \mathcal{F}^T(t).
\end{equation}

\noindent Since $\alpha_i(t)$ is the proportion of the total number of cases that $\mathcal{F}_i(t)$ represents, the condition $1=\sum_i \alpha_i(t)$ must be satisfied. Thus, using \eqref{renewal_het} with \eqref{F_prop}, we obtain:

\begin{equation}\label{renewal_het_prop}
    \mathcal{F}^T(t)= \mathcal{R}^T(t) \int_{0}^{\infty}  g^T(t,\tau)\mathcal{F}^T(t-\tau) d\tau,
\end{equation}

\noindent where $\mathcal{R}^T(t)= \bm{\alpha \cdot \overline{\mathcal{R}}}$ is the reproduction number of the system and

\begin{equation}\label{g_prop}
    g^T(t,\tau) = \frac{\sum_i \alpha_i(t) \overline{\mathcal{R}}_i(t) \; \overline{g}_i(\tau)}{\sum_i \alpha_i(t) \overline{\mathcal{R}}_i(t)} .
\end{equation}

Since $ \mathcal{R}^T (t) $ is the scalar product between $ \bm {\alpha} $ and $ \bm{\overline{\mathcal{R}}} $, we can interpret it as the projection of the $ \bm{\overline{\mathcal{R}}} $ over the fractions $ \bm {\alpha} $ forming a weighted average. Thus, it is interesting to note that we can describe the system as an average of its heterogeneities. Since the definition of the total number of reproductions $ \mathcal{R}^T $ is very general, it is not always meaningful. For example, if we form a system of two independent dynamics, that is, ($ \mathcal{R}_{ij} = 0 \text{ for } i \neq j $) it is still possible to obtain a $ \mathcal{R}^T $, even if there is no meaning to it. The $\alpha_i(t)$ functions are the key for analysing the feasibility of a total reproduction number that has a dynamical meaning.

We focus our attention on a case where the $\alpha_i(t)$ appears naturally in the equations. If $\Omega_{ij}(t,\tau)$ can be separated in a function of $t$ depending on the "$i$" index and another function of $t$ and $\tau$ depending on the "$j$" index, then it can be written as

\begin{equation}\label{Omega_prop}
    \bm{\Omega} = \bm{\alpha} \otimes \overline{\bm{\Omega}}= \Big [\alpha_i(t) \overline{\Omega}_j(t,\tau) \Big] 
\end{equation}

\noindent where $ \otimes $ represents a tensor product. We define $ \overline {\Omega}_j = \sum_i \Omega_ {ij} $, and noticed that $ \overline {A} _j = \sum_ {k} \overline {\Omega} _k \Gamma_ {kj} $ e $ \bm {A} = \bm {\alpha} \otimes \overline {\bm { A}} $. Thus, $ \bm {\Omega} $ and $ \bm {A} $ can be separated in proportions, $ \bm {\alpha} $ and $ \overline {\bm {\Omega}} $. In fact, the above equation also impacts \eqref{R_bar_j} and \eqref{J_bar} which can be factorized in terms of $ \bm {\mathcal {R}} = \bm {\alpha} \otimes \overline {\bm {\mathcal {R} }} $ e $ \bm {\mathcal {J}} = \bm {\alpha} \otimes \overline {\bm {\mathcal {J}}} $. Using \eqref{g_bar} we can also get $ g_ {ij} (t, \tau) = \overline {g} _j (t, \tau) $. Furthermore, because the next generation matrix is obtained from a tensor product of vectors, the largest eigenvalue of $ \bm {\mathcal {R}} $ corresponds to the scalar product of $ \bm {\overline {\mathcal {R }}} $ e $ \bm {\alpha} $, that is $ \mathcal {R} ^ T $. Thus, in these systems the total reproduction number assessed at the disease-free equilibrium point, $t=0$, corresponds to the basic reproduction number, $\mathcal {R} ^ T (0)= \mathcal{R}_0$. Also, in that case, the $\mathcal{R}^T$ is the spectral radius of $\bm{\mathcal{R}}$, as in \cite{van den Driessche}. This is very common in disease transmission models, in fact the SIR and SEIR model are examples of this case.

\subsection{Estimations with real data }

So far, we've developed a general framework to estimate the reproduction numbers from the rate of new infections $\mathcal{F}$. However, when we deal with real data what we have is the collection of all of the new infections in an $\Delta t$ period of time, which leads to the definition of $\bm{\mathcal{B}}$ and $\bm{\mathcal{T}}$. 

\begin{equation}\label{reported_cases}
    \mathcal{B}_i(t) = \rho_i \int_{t} ^{t+\Delta t} \mathcal{F}_i(t') dt', \qquad \mathcal{T}_{ij}(t) = \rho_i \int_{t} ^{t+\Delta t} \mathcal{J}_{ij}(t') dt'.
\end{equation}

 Therefore $\mathcal{B}_i(t)$ are the reported cases in $X_i$, where $\rho_i$ is a constant related to a higher or lower notification, due to sub/super-notification. Analogously we define $\mathcal{T}_{ij}(t)$ as the number of reported cases in $X_i$ due to previous cases in $X_j$. By assuming that $\mathcal{R}_{ij}$, $g_{ij}(t,\tau)$ and $\mathcal{F}_i $ are approximately constants during a $\Delta t$ interval, we use \eqref{J_ij_R} and \eqref{F_het} to derive

\begin{equation}\label{T}
    \mathcal{T}_{ij}(t) =  \mathcal{R}_{ij} (t) \sum_{\tau=0}^{t} \frac{\rho_i}{\rho_j} \; g_{ij} (t,\tau) \mathcal{B}_j(t-\tau)  \Delta t
\end{equation}

\noindent for $\mathcal{B}_i(t)=\sum_j^n\mathcal{T}_{ij}$. \eqref{T} is a general form for the discrete version of the renewal equation. In fact, by using \eqref{renewal_het_prop} we are able to recover a well known result in literature \cite{Fraser}

\begin{equation}
    \mathcal{R}^T(t)= \frac{\mathcal{B}^T(t)}{\sum_{\tau=0}^{t} \frac{\rho_i}{\rho_j} \; g^T (t,\tau) \mathcal{B}^T(t-\tau)\Delta t },
\end{equation}

\noindent where $\mathcal{B}^T(t) = \sum_i ^n \mathcal{B}_i(t)$.

\section{Explicit expression of $\mathcal{R}(t)$ for two meta-population models}

In this section we analyse two meta-population models detailed in Supplementary Material 2. The methodology developed on this study is applied to both models to obtain the correspondent reproduction numbers and generation interval distributions, see Supplementary Material 1 for detailed calculations.  

\subsection{SIR-type meta-population model}

In this section we will consider a meta-population model takes into account groups of spatially separated "island" populations with some level of interaction. Such models are widely used for systems in which the movement of individuals between meta-populations is considered \cite{van den Driessche_spatial,Bichara,Lloyd,Lajmanovich,Arino,Keeling,Miranda}. Since each population is connected with the others, the system can be interpreted as a network  for which the nodes represent the meta-populations and the weight of their edges represent the intensity of the movement between them. This type of model does not describe the daily movement of individuals explicitly, but as an interaction of the meta-populations. It is suitable when the population sizes are not permanently affected by the flow of individuals, as in the case of commuter movement between locations of residence, work and study. This type of movement is obligatory cyclical, predictable and recurring regularly, most of the time on a daily basis. In this model, we assumed that each meta-population $i$, with $N_i$ individuals, has its own transmission rate $\beta_i(t)$. Also, the movement of individuals between meta-populations is taken into account, where we introduce $\Phi_{ij}(t)$ as the density of flow between the populations $i$ and $j$. That is, the amount of $i$ resident individuals commuting from $i$ to $j$ divided by the total population of $i$, $N_i$. The parameters $\beta_i(t)$ and $\Phi_{ij}(t)$ are time dependent, so we can incorporate the changes in the behavior of the populations on those variables.In Supplementary Material 2 we derive this model, inspired by \cite{Miranda}, and in Supplementary Material 1 we obtain the corresponding reproduction numbers and generation interval distribution for it, which reads:

\begin{equation}
    R_{ij}(t) = S_i(t) \frac{\lambda_{ij}(t)}{\gamma}, \qquad g_{ij}(\tau)= g(\tau)= \gamma \; e^{-\gamma \, \tau}.
\end{equation}

\noindent Here $\lambda_{ij}$ is related to the transmission of the disease from a meta-population ``$i$'' to other meta-population ``$j$'' and is derived based on simple assumptions about the commuter movement of individuals in the network, see Supplementary Material 2. Notably, if we isolate the meta-populations from the network, $\Phi_{ij}(t)=0$, for all $i$ and $j$, then the reproduction numbers and the generation interval distributions becomes identical to the classical SIR model \cite{Nishiura}.

\subsection{A meta-population model for Covid-19 (SEIIR)}

The classical SIR model is a very simple and qualitative approach to disease transmission dynamics, but it does not provide the best description for most diseases. For now on, we will be focusing on a model for a specific disease, the SARS-Cov-2 coronavirus. In this case, the transmission can be facilitated by the existence of individuals whose symptoms are very weak or even nonexistent \cite{Li_R}, therefore, this heterogeneity can change the dynamics. In order to have consistence description of this aspect, it is wise to assume that the existence of two classes of infected individuals, the symptomatic and the asymptomatic/undetected ones, as considered in a general model for the SARS-Cov-2 coronavirus \cite{Oliveira}. Therefore, we now take into account infected individuals who do not need to be hospitalized and are not recorded in official registered data, thus becoming undetectable. For the sake of simplicity, we will refer to those individuals only as asymptomatic ones, for simplicity. In Supplementary Material 2 we proceed to derive this model, based in the meta-population SIR-type approach described in the previous section. In Supplementary Material 1 the expressions for the reproduction number and generation interval distribution are detailed derived. There, we show that we only need $\bm{\mathcal{R}}$ for $i,j\leq n$ to describe the dynamics. Thus, in this main framework, whenever we refer to $\bm{\mathcal{R}}$ or $\bm{g}$ we will be alluding to their $i,j\leq n$ elements. Thus, we obtain:

\begin{equation}
    R_{ij}(t) = S_i(t)\lambda_{ij}(t) \Big[ \frac{p}{\gamma_s} + \frac{\delta (1-p)}{\gamma_a}\Big], \qquad g_{ij}(\tau)= g(\tau)= \frac{\frac{p}{\gamma_s} g^s(\tau) + \frac{\delta (1-p)}{\gamma_a}g^a(\tau)}{\frac{p}{\gamma_s} + \frac{\delta (1-p)}{\gamma_a}}.
\end{equation}

\noindent the $\lambda_{ij}$'s expressions are the same presented for the SIR-type model in Supplementary Material 2. $\delta$ is a factor that reduces or enhances the asymptomatic infectivity, $p$ is the proportion of individuals that becomes symptomatic when infected, $\gamma_a$ and $\gamma_s$ are the recover rates of the asymptomatic and symptomatic individuals, respectively. $g^a(\tau)$ and $g^s(\tau)$ are expressed in terms of exponential functions described according to \eqref{eqgas}. 

\begin{equation}\label{eqgas}
      g^a(\tau) = \frac{\kappa \gamma_a}{\gamma_a - \kappa} (e^{-\kappa \tau} - e^{-\gamma_a \tau} ), \qquad \qquad \qquad g^s(\tau) =  \frac{\kappa \gamma_s}{\gamma_s - \kappa} (e^{-\kappa \tau} - e^{-\gamma_s \tau} ) .
\end{equation}

\noindent Whereby, if we isolate the meta-populations from the network, $\Phi_{ij}(t)=0$, for all $i$ and $j$,  the reproduction numbers and generation interval distributions returns to the expression obtained in \cite{Oliveira}.

\section{Applications for the meta-population models using actual data}

In this section, we present results for the methodology applied to the meta-population model developed in the previews section. We use actual data of the first six months of the COVID-19 pandemic in Brazilian cities, such as: reported cases in each municipality, daily commuter movement due to work between municipalities and daily mobility tends towards workplaces. In Supplementary Material 3 we derive and present the expressions and parameters needed to estimate the reproduction numbers for both meta-population models. Thus, we obtain a daily time series of the reproduction numbers for each model.

\subsection{Database}
In this work, we use daily notifications of new cases due to COVID-19 in Brazil, obtained from public websites: \url{https://covid.saude.gov.br/} and \url{https://brasil.io/datasets/} , which provide data from the Health Ministry. We obtained intermunicipal commuter movement of workers and students data from a study on population arrangements and urban concentrations in Brazil conducted by IBGE (Brasilian Institute of Geography and Statistics) in 2015 that can be found at \cite{ibge}. In addition, we obtain daily mobility data for each Brazilian state from a public report by Google, accessed at: \url{https://google.com/covid19/mobility/}. 

 We used the data observed until September 14th, 2020, and performed a 10-day moving average, in order to attenuate noise and better express the data trend. To take the social distancing restrictions into account, we considered only the commuter movement data related to work, since, with the mitigation measures of COVID-19, the flow due to education were significantly reduced. In addition, because the movement towards work also dropped due to social isolation policies, we used the mobility data obtained from the community mobility report provided by Google to estimate this reduction. This database compares for each state the daily mobility  to workplaces with the past trends, therefore, we can access the percentage of reduction in commuting to work. A moving average is performed in the mobility time series and we reduce the intermunicipal work flow according to the percentage indicated on the data. Thus, we obtain the flow of individuals due to commuter work that leads to $\Phi_{ij}(t)$. The parameters used to feed the model were obtained in \cite {Jorge} and are displayed in Supplementary Material 3.

The data from the state of Rio de Janeiro was selected for this analysis. Ten cities with the highest commuter flow with the capital Rio de Janeiro (RJ), were chosen, namely: Duque de Caxias (DdC), Nova Iguaçu (NI), São João de Meriti (SJdM), Niterói (Nt), São Gonçalo (SG), Belford Roxo (BR), Nilópolis (Ns), Mesquita (Mq), Queimados (Q), Magé (Ma). All of the chosen cities for this work are part of the Rio de Janeiro's metropolitan region. Additional information about the municipalities is presented in Supplementary Material 4.

\subsection{Analyses of the results}

\begin{figure}
     \centering

    \includegraphics[width={1.\linewidth}]{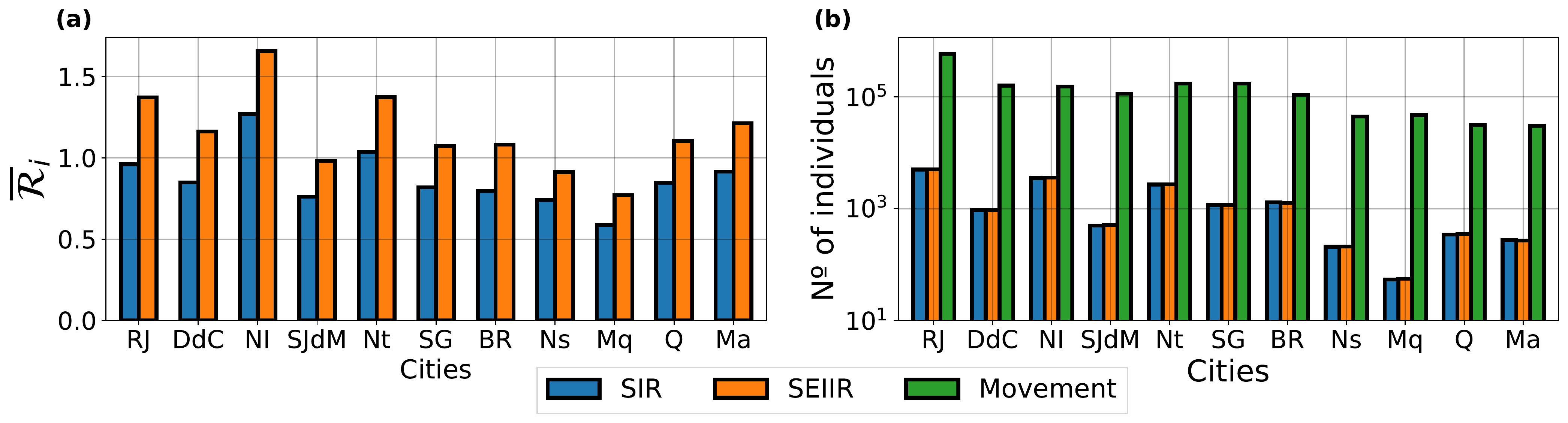}

    \caption{\textbf{Comparison between SIR and SEIIR outputs.} SIR results in blue and SEIIR ones in orange. The bar graph (a) compares the SIR and SEIIR $\overline{\mathcal{R}}_i$ averages in time for all municipalities. In (b), for each municipality, estimations for the total number of exported cases that are reported for both models are displaced with the total commuter movement in that city. The names of the municipalities are abbreviated using acronyms: Rio de Janeiro (RJ), Duque de Caxias (DdC), Nova Iguaçu (NI), São João de Meriti (SJdM), Niterói (Nt), São Gonçalo (SG), Belford Roxo (BR), Nilópolis (Ns), Mesquita (Mq), Queimados (Q), Magé (Ma).}
        \label{fig:1}
\end{figure}

 In our first results, shown in Figure \ref{fig:1}, we present a comparison between the SIR and SEIIR outputs. Using the daily time series of the reproduction numbers, see Supplementary Material 3, we obtain the series of $\overline{\bm{\mathcal{R}}}(t)$ and $\bm{\mathcal{T}}(t)$ from \eqref{R_bar_j} and \eqref{T}, respectively. We observe that the $\overline{\bm{\mathcal{R}}}$ of SEIIR model is, in the average, $33\%$ higher then the corresponding results for the SIR model. On the other hand, the estimations of the total number of exported cases that are reported, $ \sum _{t}\sum_{i} ^n \mathcal{T}_{ij}(t)$ for $i \neq j$, is very similar for both SIR and SEIIR models Also, it seems that the total commuter movement, which is the sum of all the inflow and outflow happening in a municipality, is not the only main factor that determine the number of exported cases of a municipality. This non-linearity can be observed when comparing São João de Meriti (SJdM) and Niterói (Nt) or Duque de Caxias (DdC) and Nova Iguaçu (NI) (see Figure \ref{fig:1}b). Those municipalities have a similar amount of total flow but very different results for the exported cases. Interestingly, even not having the highest $\overline{\mathcal{R}}_j$, the capital, Rio de Janeiro (RJ) presents the largest amount of exported cases, which also showcases the non-linear dynamics of the phenomenon.

\begin{figure}
     \centering

    \includegraphics[width={0.75\linewidth}]{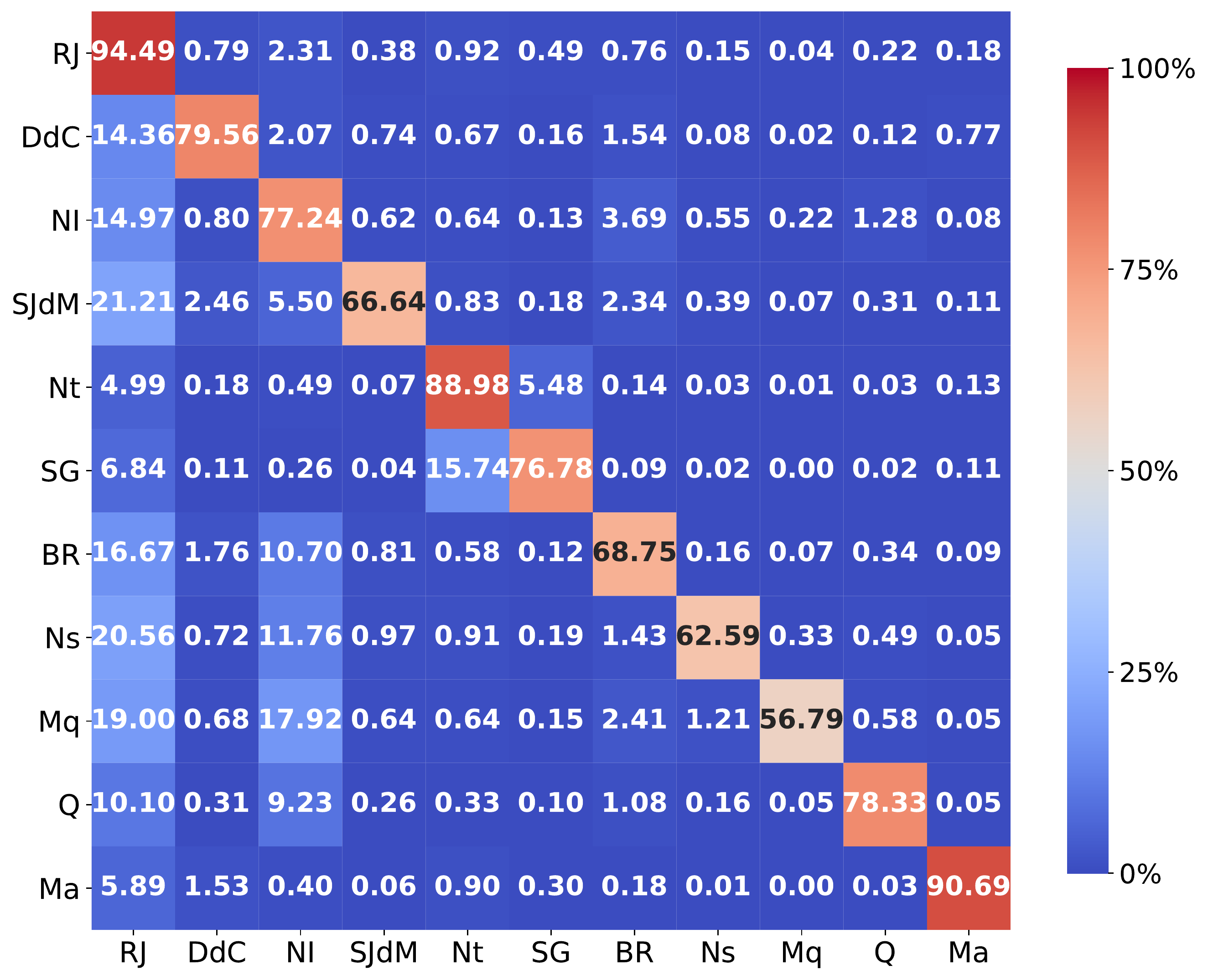}

    \caption{\textbf{Influence on number of cases of a municipality in another.} The heatmap captures the average, through time, influence on number of cases of a municipality in another as a proportion of the total number of daily cases, $\mathcal{T}_{ij}/\mathcal{B}_i$. $i$ corresponds to the rows and $j$ to the columns. }
        \label{fig:2}
\end{figure}

From now on, we will only look at the results of the SEIIR model. With $\mathcal{T}_{ij}(t)$ we are able to access the contribution of each municipality on the outbreaks happening in the state. Thus, by dividing $\mathcal{T}_{ij}(t)$ by $\mathcal{B}_i(t)$ in every time step, we obtain a time series for the proportion of the total cases in ``$i$'' generated by ``$j$''. Therefore, we proceed into evaluating the mean value through time of $\mathcal{T}_{ij}/\mathcal{B}_i$ , where it is displaced in Figure \ref{fig:2}. It is observed a high autochthonous behavior on the disease transmission, whereby the highest influence of a city is on itself. Therefore, most of the cases generated in a municipality are caused by its own individuals. However, we also identify cities where the cases generated by other municipalities on it are very important. The $\bm{\mathcal{R}}(t)$ matrix also corroborates the presence of an important autochthonous behavior as its diagonal elements correspond to the highest values of the reproduction numbers. We also observed lots of very smalloff-diagonal elements low values through out the matrix.

\begin{figure}
     \centering

    \includegraphics[width={0.63\linewidth}]{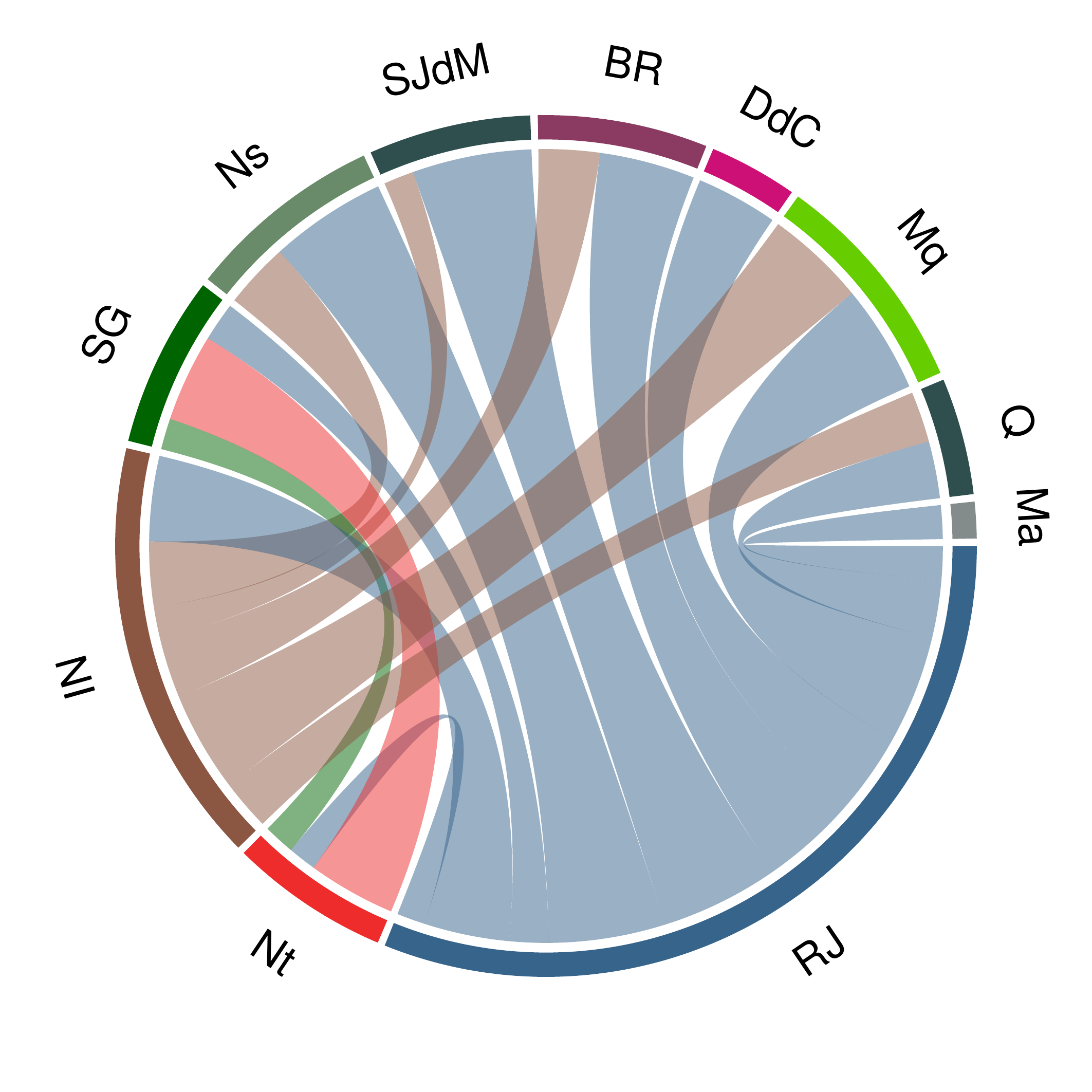}

    \caption{\textbf{Visualization of the influence between municipalities.} Here, we present a visualization of the results from Figure \ref{fig:2}. Only non-autochthonous influences above 5\% were considered. The thickness of the lines connecting municipalities is proportional to the number of cases that one generates on the other. The color of each line represents the municipality that is generating the cases.}   
        \label{fig:3}
\end{figure}

In Figure \ref{fig:3} we illustrate the results of Figure \ref{fig:2}, whereby only non-autochthonous influences above 5\% are considered. Thus, we must highlight the capital, Rio de Janeiro (RJ), as the most important agent on the disease transmission to the municipalities on the network. However, cities like Nova Iguaçu (NI) also presents itself as a relevant disseminator of the pathogen. As shown in Figure \ref{fig:1}, Nova Iguaçu NI is the highest city in case exportation, besides the capital. In Figure \ref{fig:3} we identify cities like: São João de Meriti (SJdM), Belford Roxo (BR), Nilópolis (Ns), Mesquita (Mq) and Queimados(Q) as the main receptors of those cases. Niterói (Nt), even having a NI-like number of exported cases, did not presented a high influence on a lot of cities. On the other hand, Niterói (Nt) generates a significant amount of cases in São Gonçalo (SG), highlighting the importance of the connection of those two municipalities.

\begin{figure}
     \centering

    \includegraphics[width={1.\linewidth}]{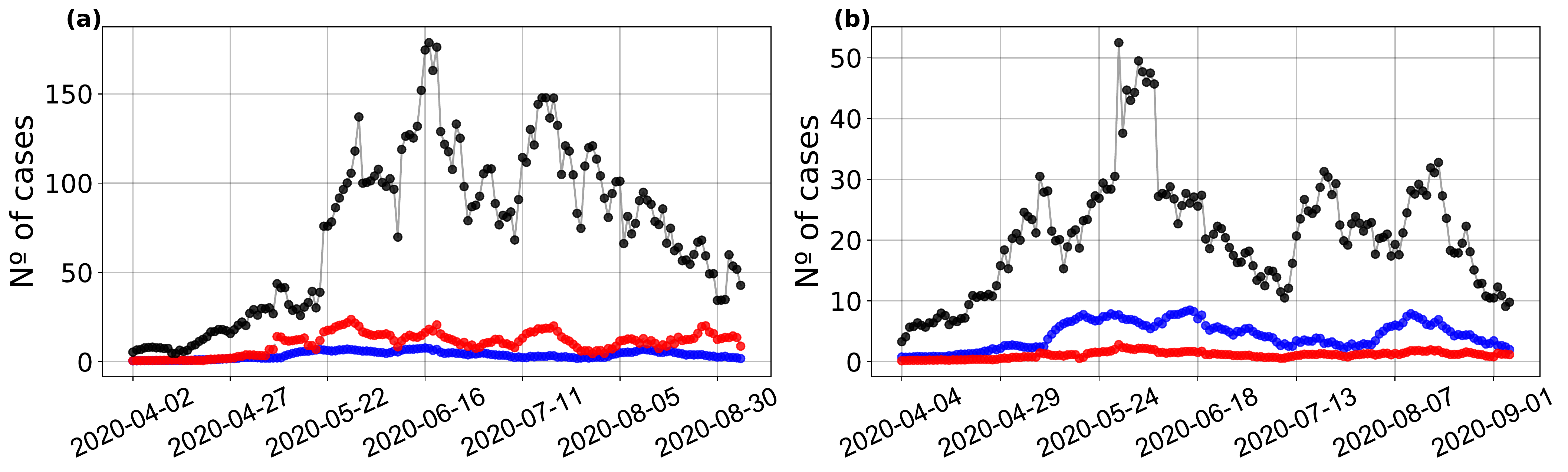}

    \caption{\textbf{Reconstruction of the time series.} The black dots represents the daily reported cases in (a) São Gonçalo (SG), (b) São João de Meriti (SJdM). The blue dots are the amount of cases generated in (a) São Gonçalo (SG), (b) São João de Meriti (SJdM) due to the capital, Rio de Janeiro (RJ). In red we have the number of cases reported in (a) São Gonçalo (SG) due to Niterói (Nt); (b) São João de Meriti (SJdM) due to Nova Iguaçu (NI).}   
        \label{fig:4}
\end{figure}

We illustrate the time series reconstruction in Figure \ref{fig:4} where two scenarios are displaced. In the first one, we focus on the cases reported in São Gonçalo (SG) and compare total amount, $\mathcal{B}_i(t)$,  with the number of daily cases in SG generated by Rio de Janeiro (RJ) and Niterói (Nt). We observe that Nt has a predominance over the capital, presenting a higher number of cases generated in all times. The second scenario is related to the total number of cases in São João de Meriti (SJdM) and the contributions due to Rio de Janeiro (RJ) and Nova Iguaçu (NI). In this case, the capital presents the largest number of cases generated in that city, besides the city itself. It is also interesting to observe in both scenarios how the cities contributions merge into a part of the total cases notified on a municipality.

\section{Discussion}

Understanding and dealing with infectious diseases is an on going challenge to humankind. The methodology presented in this work is very general and can be applied to multiple disease transmission systems. By merging the model dynamics with the available epidemic data, this method is able to estimate key epidemiological factors, like the reproduction numbers and the generation interval distributions. These theoretical results are the basis for a lot of possible data analyses, specially for highly heterogeneous systems that have been demanding for a suitable methodology for estimating the reproduction number through actual data. The method is robust and reproduces known results in literature, as shown in Supplementary Material 1 for the SIR, SEIR and SEIIR models. This methodology opens room for the analyses of more sophisticated models, that leads to a better understanding and control of infectious diseases.

With the emerging of the COVID-19 pandemic, due to SARS-Cov-2 coronavirus, in December 2019, scientists allover the world have joined forces to formulate control strategies \cite{Manica}. Factors like the presence of asymptomatic individuals, risk groups and spatial distribution of the disease brings out a system with multiples layers of heterogeneity \cite{Li_R, Eikenberry, Miranda, Gomes}. 
Although the application of some vaccines starts recently, mitigating COVID-19 with  non-pharmacological strategies is essential. Therefore, the second half of this work we focused on the application of the developed methodology for two models with spatial heterogeneity, one focusing on the COVID-19 specific dynamics. Each municipal demography, captured by $\beta$ parameters, and the commuter movement were combined by the model lead to some interesting results. The reproduction numbers are obtained for both models indicate that they only differ by multiplying factors, as $1/\gamma$ and $ \frac{p}{\gamma_s} + \frac{\delta (1-p)}{\gamma_a}$, as in \cite{Oliveira, Jorge}. The measured results shows that the asymptomatic presence is related to an increase on the reproduction number, because the model predicts more infected individuals than what is reported. Another substantial result is obtained when comparing the exported cases with the total commuter movement in a municipality. In this case, the relationship is not direct, since the intricate topology of the network fosters a non-linear dynamic in the phenomenon. This result points up the importance of a model to provide epidemiological meaning of the available data. For this specific case, it became evident that analysing the system only by the commuter movement data wouldn't be enough to point out the key cities on the disease transmission dynamic, as we did using the models and methodology. 

The results of the model display the role of each municipality on the epidemic on the network. Cities like Nova Iguaçu, Niterói and São Gonçalo pops out as important agents on the spread of the pathogen through out the metropolitan region of Rio de Janeiro, Brazil.
The capital is highlighted as the main hub of spreading, which is related to its high incidence of the disease combined with its central role on the movement behavior of individuals on the network, as also observed in \cite{Jorge, Miranda, Liu}. However, cases like Niterói and São Gonçalo, where the interaction of both cities is higher than the capital's influence, cannot be neglected. This call's the attention for the relevance of analysis, like in this work, that provides the epidemiological interpretation of data. In this work, we choose to present an analysis with actual data in which the reproduction number is not the main result. We proceeded with this approach to portrait the reproduction number not as a dead end result, but as a tool for obtaining deeper analyses, such as the reconstruction of the time series and the number of exported cases.      

Our results for the Rio de Janeiro metropolitan area, however, have limitations. The model developed in this work provides a very simple approximation of intercity commuter flow, while more sophisticated models available in the literature are required to provide more precise description of the movement behavior. Another limitation of this work is that it does not consider the interstate flow, since the state of Rio de Janeiro was portrayed as a closed system Also, we did not take into account further heterogeneous features, which could be accounted as well by the general theoretical formalism, besides space and asymptomatic presence, leaving a gap for other key COVID-19 dynamics characteristics, like age-groups. In addition, the examples presented in this work considered actual reported data of confirmed cases and mobility, which may present problems of underestimation and report delay. Taking those limitations into account, the results displaced here are still able to give substantial understanding of the system studied, which is a common feature in mathematical modelling. Finally, we reinforce the importance of the method proposed on this work and highlight its broad application on infectious disease models.

\enlargethispage{20pt}

\ethics{Since all data handled in this study is publicly available, an approval by an ethics committee is not required, according to Resolutions 466/2012 and 510/2016 (article 1, sections III and V) from the National Health Council (CNS), Brazil.}

\dataccess{All data and codes are gathered and presented in our public GitHub repository at \url{https://github.com/danielcpj/Rt-heterogeneous-models}}


\aucontribute{DCPJ, STRP and RFSA conceived of and designed the methodology presented in this study. DCPJ, STRP, JGVM and JFO formulated and interpreted the meta-population models and applications. DCPJ performed the data analysis and drafted the manuscript. All authors read, reviewed and approved the manuscript.}

\competing{The authors declare that they have no competing interests.}

\funding{DCPJ was funded by a Scientfiic Initiation scholarship from CNPq (process
number 117568/2019-8). JFO was supported by the Fiocruz Program of Promotion of Innovation - innovative ideas and products - COVID-19, orders and strategies - Inova Fiocruz (Processo VPPIS-005-FIO-20-2-40), and the Center of Data and Knowledge Integration for Health (CIDACS) through the Zika Platform - a long-term surveillance platform for Zika virus and microcephaly (Unified Health System (SUS), Brazilian Ministry of Health). STRP was supported by an International Cooperation grant (process number INT0002/2016) from Bahia
Research Foundation (FAPESB). RFSA was supported by Brazilian agency CNPq through Grants No. 422561/2018-5 and
304257/2019-2. STRP and RFSA were supported by the National Institute of Science and Technology - Complex Systems from CNPq, Brazil. JGVM acknowledges the support of the National Council of Technological and Scientific Development, CNPq, Brazil (Grant number: 307828/2018-2).}

\ack{The authors acknowledge the discussions and suggestions from members of the CoVida Network (\url{http://www.redecovida.org}).}

\beginsupplement 
\section*{Supplementary Materials}
\paragraph*{Supplementary Material 1-}{ Methodology applied for epidemiological compartment models}
\vspace{-10pt}
\paragraph*{Supplementary Material 2-}{ Meta-population models formulation.}
\vspace{-10pt}
\paragraph*{Supplementary Material 3-}{Expressions and parameter values to estimate the reproduction numbers of the meta-population models}
\vspace{-10pt}
\paragraph*{Supplementary Material 4-}{Additional information about the municipalities.}


\end{document}


\flushbottom
\maketitle
\thispagestyle{empty}

\beginsupplement 

\newpage
\topskip0pt
\vspace*{\fill}
\begin{center}
\begin{minipage}{.6\textwidth}
\Large{\textbf{Supplementary Material 1}}\\
\large{\textcolor{gray}{\textbf{Methodology applied to epidemiological compartment models}}}
\end{minipage}
\end{center}
\vspace*{\fill}

\newpage

\section*{Applying the method to epidemiological models}

    We proceed to illustrate the application of the method developed in this work to compartmental epidemic models. We chose two simple models that are known in literature and two variations of a meta-population model that is used in the main framework and developed in Supplementary Material 2. We seek to show the method step by step, presenting its detailed calculations. All models presented are composed of ordinary differential equations. In the transitions between infected compartments described in these models, $ \mathbfcal{V} (t, \tau) $, the the infected compartments $\bm {x}(t,\tau)$ appear with linear dependence. Therefore we can simplify: 

\begin{equation}\label{cte}
    \dfrac{\partial\overline{\mathcal{V}_i}}{\partial u_j} = \dfrac{\partial \mathcal{V}_i}{\partial x_j}=cte \qquad \text{and} \qquad \frac{d}{d\omega} \bm{u}(\omega) =  -\frac{\partial {\mathbfcal{V}}}{\partial \bm{x}} \bm{u}(\omega).
\end{equation}

\noindent In addition, all the parameters of the models in this Supplementary Material are constant, which leads to

\begin{equation}
    \bm{\Omega}(t)= \frac{\partial}{\partial \bm{X}} {\mathbfcal{F}}(t) \qquad \text{and} \qquad \overline{\bm{\Gamma}}(\omega) =  \bm{\Gamma}(\tau)
\end{equation}

\subsection*{SEIR model}

The SEIR model is designed by introducing a new exposed $ E $ stage of the disease into the SIR model \cite{SIRfirst}. We can assume that when an individual becomes infected, it must go through a latency period before showing its first symptoms and starting to infect other individuals. This is accomplished by introducing the exposed compartment $ E $ and its removal rate $ \kappa $. Thus, all individuals who are infected start in the exposed state and, on average, after a $ 1 / \kappa $ latency time are introduced into the $ I $ compartment. It is also possible to introduce a factor related to a $ \epsilon $ pre-symptomatic infection. This way, individuals can start to infect in the exposed compartment, before presenting symptoms. Thus, this model, considering pre-symptomatic infection, can be written as

 \begin{align}
\frac{dS}{dt}& =-   \frac{\beta S}{N}\Big [  I + \epsilon E \Big ],\\
\frac{dE}{dt}& = \frac{\beta S}{N}\Big [  I + \epsilon E \Big ] - \kappa E,\\
 \frac{dI}{dt}& = \kappa E - \gamma I,\\
\frac{dR}{dt}&= \gamma I.
\end{align}

Thus, we can sort the infected compartments as $\bm{X}(t)=[E(t), \;I(t)]$. In this way the distributions of the infectious phase are defined as $\bm{x}(t,\tau)=[i_e(t,\tau) ,i_i(t,\tau)]$. We get
 ${\mathbfcal{F}}(t)$ e ${\mathbfcal{V}}(t,\tau)$:

\begin{equation}
      {\mathbfcal{F}}(t) = 
      \begin{pmatrix}
  \frac{\beta S}{N}\big [  I + \epsilon E \big ]\\[1ex] 
    0
\end{pmatrix}, \qquad 
         {\mathbfcal{V}}(t,\tau) =  \begin{pmatrix}
  \kappa i_e(t,\tau)\\[1ex] 
    \gamma i_i(t,\tau) - \kappa i_e(t,\tau)
\end{pmatrix},
\end{equation}

\noindent as we recover the sub-set of equations for the infected compartments with:

\begin{equation}
    \frac{d}{dt} \bm{X}(t) = \mathbfcal{F}(t) - \int_0 ^{\infty} \mathbfcal{V}(t,\tau) d\tau.
\end{equation}

\noindent The change from $ E $ to $ I $ is not considered a new infection, but the progression of the disease stage. Therefore, new infections only occur in the exposed compartment $ E $, causing $\mathcal{F}_2(t)=0$. Thus, from ${\mathbfcal{F}}(t)$, we obtain $\bm{\Omega}(t)$:

\begin{equation}
          \bm{\Omega}(t) =  \Big[ \frac{\partial}{\partial \bm{X}} {\mathbfcal{F}}(t) \Big] =
      \begin{pmatrix}
  \epsilon \frac{\beta S}{N} & \frac{\beta S}{N} \\
    0 & 0
\end{pmatrix}.
\end{equation}

\noindent We proceed to obtain $\Gamma(\tau)$ by solving

\begin{equation}
          \frac{d}{d\omega} \bm{u}(\omega) = - \frac{\partial {\mathbfcal{V}}}{\partial \bm{x}} \bm{u}(\omega)  =
      \begin{bmatrix}
  -\kappa & 0 \\
    \kappa & -\gamma
\end{bmatrix} \bm{u}(\omega).
\end{equation}

\noindent In order to obtain

\begin{equation}
          \bm{u}(\omega) = 
           \begin{bmatrix}
   e^{-\kappa\omega} & 0 \\
     \frac{\kappa}{\gamma -\kappa} [e^{-\kappa\omega} - e^{-\gamma\omega} ]& e^{-\gamma\omega}
\end{bmatrix}    \bm{u}(0).
\end{equation}

\noindent Therefore:

\begin{equation}
          \bm{\Gamma}(\tau) =                \begin{bmatrix}
   e^{-\kappa\omega} & 0 \\
     \frac{\kappa}{\gamma -\kappa} [e^{-\kappa\tau} - e^{-\gamma\tau} ]& e^{-\gamma\tau}
\end{bmatrix}
\end{equation}

With $\bm{\Omega}(t)$ and $\bm{\Gamma}(\tau)$
we perform the multiplication between the matrices, in order to obtain

\begin{equation}
          \bm{A}(t,\tau) =
          \begin{bmatrix}
  \epsilon \frac{\beta S}{N} & \frac{\beta S}{N} \\
    0 & 0
\end{bmatrix}
\begin{bmatrix}
   e^{-\kappa\tau} & 0 \\
     \frac{\kappa}{\gamma -\kappa} [e^{-\kappa\tau} - e^{-\gamma\tau} ]& e^{-\gamma\tau}
\end{bmatrix}.  
\end{equation}

\noindent All terms in the second line are null, leaving only

\begin{align}
    A_{11}(t,\tau)&=  \epsilon \frac{\beta S}{N}e^{-\kappa\tau} +  \frac{\beta S}{N}\frac{\kappa}{\gamma -\kappa} [e^{-\kappa\tau} - e^{-\gamma\tau} ],\\
    A_{12}(t,\tau)&= \frac{\beta S}{N} e^{-\gamma\tau}.
\end{align}

\noindent Performing the integral from zero to infinity with respect to $ \tau $ we obtain the reproduction numbers of the system
${\mathbfcal{R}}(t)$:

\begin{equation}
    {\mathbfcal{R}}(t)=\beta\frac{S}{N}
    \begin{pmatrix}
    \dfrac{\epsilon}{\kappa}+ \dfrac{1}{\gamma}  &  \dfrac{1}{\gamma} \\[2ex]
    0 & 0
\end{pmatrix}.
\end{equation}

\noindent Where it is clear that

\begin{equation}
    \bm{\overline{\mathbfcal{R}}}=  \beta\frac{S}{N}  \begin{pmatrix}
    \dfrac{\epsilon}{\kappa}+ \dfrac{1}{\gamma}  \\[2ex] \dfrac{1}{\gamma} 
\end{pmatrix}.
\end{equation}

Since there is only generation of new infected in the exposed compartment, we have to $ \mathcal {F} _1 (t) = \mathcal {F} ^ T (t) $. Thus, the vector $ \bm {\alpha} (t) $ can be written as $ \bm {\alpha} = \big (1 , 0 \big) $, thus

\begin{equation}
{\mathbfcal{R}} = \bm{\alpha} \otimes \bm{\overline{\mathbfcal{R}}}=
    \begin{pmatrix}
    1 \\
    0
\end{pmatrix} \bigotimes
\beta\frac{S}{N}
    \begin{pmatrix}
    \dfrac{\epsilon}{\kappa}+ \dfrac{1}{\gamma}  \\[2ex] \dfrac{1}{\gamma} 
\end{pmatrix}.
\end{equation}

\noindent Thus, to obtain the total reproduction number of the system, it is enough to make the scalar product between $\bm{\overline{\mathbfcal{R}}}$ and $\bm{\alpha}$:

\begin{equation}
    \mathcal{R}^T(t)= \bm{\alpha} \bm{\cdot} \bm{\overline{\mathbfcal{R}}} = \frac{\beta S}{N} \Big [ \frac{\epsilon}{\kappa}+ \frac{1}{\gamma} \Big ],  
\end{equation}

\noindent which leads to

\begin{equation}
    g^T(\tau)=   \frac{\epsilon \; e^{-\kappa\tau} +  \frac{\kappa}{\gamma -\kappa} [e^{-\kappa\tau} - e^{-\gamma\tau} ]}{\epsilon/\kappa+ 1/\gamma} . 
\end{equation}

When we make $ \epsilon \to 0 $, we retrieve the result of the reproduction number of the classic SEIR model (without pre-symptomatic infection),$ \mathcal {R} _0 = \beta / \gamma $. Similarly, the generation interval distribution of the SEIR model is also reduced to the well-known form in the literature \cite{champredon2018equivalence}, demonstrating robustness in the method. It is trivial to apply the method to the SIR model, whose analysis corresponds to tanking a limit of $\kappa \to \infty$. Therefore, both the reproduction number and the generation interval distribution, $g(\tau)= \gamma e^{-\gamma \tau}$, return to the known results from literature \cite{nishiura2009effective}.

\subsection*{SIIR model}

 This is an adaptation of the SIR model, where we include two different manifestations of the disease, $ I_1 $ and $ I_2 $. In this model, there is only one susceptible population whose individuals can evolve into two compartments that carry the infectious agent. Both types of the disease carry the same pathogen, so that individuals infected with either type can go for $ I_1 $ and $ I_2 $. When infected, an individual has the probability $ p $ of manifesting the type $ I_1 $ of the disease and $ q = (1-p) $ of manifesting the type $ I_2 $. We assume two independent infection rates $ \beta_1 $ and $ \beta_2 $ for $ I_1 $ and $ I_2 $. Likewise, each slot has its own $ \gamma_1 $ or $ \gamma_2 $ removal rate. So, we write the model equations:

 \begin{align}
\frac{dS}{dt}& =- \frac{(\beta_1 I_1 +\beta_2 I_2)S}{N},\\
 \frac{dI_1}{dt}& = p\frac{(\beta_1 I_1 +\beta_2 I_2)S}{N}  - \gamma_1 I_1,\\
  \frac{dI_2}{dt}& = q\frac{(\beta_1 I_1 +\beta_2 I_2)S}{N}  - \gamma_2 I_2,\\
\frac{dR}{dt}&= \gamma_1 I_1 + \gamma_2 I_2.
\end{align}

We sort the infected compartments as $\bm{X}(t)=[I_1(t), \;I_2(t)]$ and $\bm{x}(t,\tau)=[i_1(t,\tau) ,i_2(t,\tau)]$, where it is clear that $\bm{X}(t)= \int_{0}^{\infty}\bm{x}(t,\tau) d\tau$. We obtain ${\mathbfcal{F}}(t)$ and ${\mathbfcal{V}}(t,\tau)$ as:

\begin{equation}
      {\mathbfcal{F}}(t) =\dfrac{S}{N}
      \begin{pmatrix}
  p  (\beta_1 I_1 +\beta_2 I_2)\\[1ex]
   q (\beta_1 I_1 +\beta_2 I_2)
\end{pmatrix}, \qquad 
         {\mathbfcal{V}}(t,\tau) =  \begin{pmatrix}
  \gamma_1 i_1(t,\tau)\\
    \gamma_2 i_2(t,\tau)
\end{pmatrix}.
\end{equation}

\noindent Thus

\begin{equation}
          \bm{\Omega}(t) = \dfrac{S}{N}
      \begin{pmatrix}
    p\beta_1  & p\beta_2   \\[1ex]
    q\beta_1 & q\beta_2 
\end{pmatrix}, \qquad -\frac{\partial {\mathbfcal{V}}}{\partial \bm{x}} = 
      \begin{pmatrix}
   -\gamma_1 & 0  \\
    0 & -\gamma_2 
\end{pmatrix}.
\end{equation}

The linear O.D.E system described by the matrix $-\partial {\mathbfcal{V}}/\partial \bm{x}$ it is simple to solve, since this is a diagonal matrix. So, we get

\begin{equation}
    \bm{\Gamma}(\tau) =
    \begin{pmatrix}
    e^{-\gamma_1 \tau} & 0 \\
    0 & e^{-\gamma_2 \tau}
    \end{pmatrix},
\end{equation}

\noindent which, when multiplied by the matrix $ \bm {\Omega} (t) $ results in

\begin{equation}
              \bm{A}(t,\tau) = \dfrac{S}{N}
      \begin{pmatrix}
    p\beta_1  \; e^{-\gamma_1 \tau} & p\beta_2\; e^{-\gamma_2\tau}  \\[2ex]
    q\beta_1  \; e^{-\gamma_1 \tau} & q\beta_2 \; e^{-\gamma_2 \tau}
\end{pmatrix}
\end{equation}

\noindent It is interesting to realize that this is one of the cases where $A_{ij}(t,\tau)= \alpha_{i} \overline{A}_j(t,\tau)$, recalling that $\overline{A}_j(t,\tau)= \sum_{i} A_{ij}(t,\tau)$. Therefore $\bm{\alpha} =\big(p,q\big)$ and we can factorize the matrix into $\overline{\bm{A}}$ and $\bm{\alpha}$ as:

\begin{equation}
              \bm{A} = \bm{\alpha} \otimes \overline{\bm{A}} =        
\begin{pmatrix}
    p \\
    q
\end{pmatrix} \bigotimes
\begin{pmatrix}
    \dfrac{\beta_1 S}{N} \; e^{-\gamma_1 \tau} \\[2ex]
    \dfrac{\beta_2 S}{N}\; e^{-\gamma_2\tau}  \\
\end{pmatrix}.
\end{equation}

\noindent Where $\otimes$ represents the tensorial product, $\bm{\alpha} \otimes \bm{\overline{\bm A}}= \Big [ \alpha_i \overline{A}_j \Big ]$. Integrating $\overline{\bm{A}}(t,\tau)$ in relation to $\tau$ we get:

\begin{equation}
              \bm{\overline{\mathbfcal{R}}}(t) = \int_0 ^{\infty}\overline{\bm{A}}(t,\tau) d\tau =\dfrac{ S}{N}
\begin{pmatrix}
     \beta_1/\gamma_1\\[1ex]
    \beta_2/\gamma_2 \\
\end{pmatrix}.
\end{equation}

\noindent Therefore, we proceed to the scalar product between $\bm \alpha$ and $\overline{\mathcal{\bm{R}}}$: 

\begin{equation}
    \mathcal{R}^T (t) = \bm{ \alpha \cdot \overline{\mathbfcal{R}}}=\frac{S(t)}{N} \bigg [ p \frac{\beta_1}{\gamma_1} + q\frac{\beta_2}{\gamma_2}  \bigg ].
\end{equation}

\noindent When $t \to 0$, $\mathcal{R}_0 = p \beta_1/\gamma_1 + q\beta_2/\gamma_2 $. We realize that the total reproduction number is the sum of the reproduction numbers of the two types of infection times the percentage of occurrence of each. Thus, we proceed to obtain the distribution of the generation interval, which takes the form:

\begin{equation}
    g^T(\tau) = \dfrac{ p \beta_1 \; e^{-\gamma_1 \tau} + q \beta_2 \; e^{-\gamma_2 \tau} }{ \beta_1/\gamma_1 + \beta_2/\gamma_2 }.
\end{equation}

\subsection*{SIR-type meta-population model}

In this section we consider a meta-population model that will be used in the main framework and is detailed at Supplementary Material 2. Here we summarize the model. We consider the existence of ``$ n $'' meta-populations with coupled SIR-type dynamics. Where, due to the movement of individuals between meta-populations, an infected compartment in one meta-population can influence the disease transmission process of all the others. The coupling of the equations happens by the transmission rates $ \lambda_ {ik} $ related to the contamination process that emerges from the flow of individuals. The SIR-type model for $n$ meta-populations can be written as

\begin{align}
\frac{dS_i}{dt} =& - \sum_{j}^{n} \lambda_{ij}(t)\;I_j(t) \; S_i(t),\label{eqS}\\
\frac{dI_i}{dt}=& \sum_{j}^{n} \lambda_{ij}(t)\;I_j(t)\; S_i(t) -\gamma \; I_i(t),\label{eqI}\\
\frac{dR_i}{dt}=& \gamma I_i(t),\label{eqR}
\end{align}

\noindent in which $S_i(t)$, $I_i(t)$ and $R_i(t)$ correspond to the susceptible, infected and removed individuals that belong to the meta-population ``$i$''. The $\lambda_{ij}$ parameter is related to the transmission between the meta-populations ``$i$'' and ``$j$'', see Supplementary Material 2. We assume that the recover rate $\gamma$ is uniform for all meta-populations.The infected compartments are sorted as $\bm{X}(t)=[I_1(t), \;I_2(t), \hdots , I_n(t)]$ and $\bm{x}(t,\tau)=[i_1(t,\tau), \;i_2(t,\tau), \hdots , i_n(t,\tau)]$, where it is clear that $\bm{X}(t)= \int_{0}^{\infty}\bm{x}(t,\tau) d\tau$. We proceed to identify $\mathcal{F}_i(t)$ and $\mathcal{V}_i(t,\tau)$ as

\begin{equation}
    \bm{\mathbfcal{F}}(t) = \Bigg [  \sum_j ^n \lambda_{ij}(t) I_j(t) S_i(t)  \Bigg ], \qquad \bm{\mathbfcal{V}}(t,\tau) = \Bigg [ -\gamma \; i_i(t,\tau) \Bigg ]
\end{equation}

Therefore we get:

\begin{equation}
          \bm{\Omega}(t) = 
      \begin{pmatrix}
     \lambda_{11} S_1 &  \lambda_{12} S_1 & \ldots & \lambda_{1n} S_1   \\
    \lambda_{21} S_2 &  \lambda_{22} S_2 & \ldots & \lambda_{2n} S_2  \\
    \vdots & \vdots & \ddots & \vdots  \\
    \lambda_{n1} S_n &  \lambda_{n2} S_n & \ldots & \lambda_{nn} S_n 
\end{pmatrix},\qquad
-\frac{\partial \bm{\mathbfcal{V}}}{\partial \bm{x}} = 
      \begin{pmatrix}
     \gamma &  0 & \ldots & 0   \\
        0    &  \gamma & \ldots & 0  \\
    \vdots & \vdots & \ddots & \vdots  \\
    0 &  0 & \ldots & \gamma 
\end{pmatrix}.
\end{equation}

It is clear to see that $-\partial \bm{\mathbfcal{V}}/\partial \bm{x}$ is a diagonal matrix, such that $\dfrac{d}{d\omega}\bm{u}(\omega) = -\dfrac{\partial \bm{\mathbfcal{V}}}{\partial \bm{x}} \bm{u}(\omega)$ are of trivial solution

\begin{equation}
    \bm{\Gamma}(\tau) = e^{-\gamma \tau}  \mathds{I},
\end{equation}

\noindent where $\mathds{I}$ represents the identity matrix of dimension ``$n$''. Thus, when multiplying the matrices $ \bm {\Omega} $ and $ \bm {\Gamma} $ we arrive at:

\begin{equation}
    \bm{A}(t,\tau) = \Big [ \lambda_{ik} S_i \;  e^{-\gamma \tau}\Big ]
\end{equation}

\noindent Integramos $\bm{A}$ de forma a obter a matriz de próxima geração do sistema $\bm{\mathbfcal{R}}$, dada por:

\begin{equation}
    \bm{\mathbfcal{R}}(t) = \Bigg [ \dfrac{\lambda_{ik} S_i}{\gamma} \Bigg ].
\end{equation}

\noindent That leads to:

\begin{equation}
 g_{ij}(\tau)= g(\tau)= \gamma \; e^{-\gamma \, \tau}.
\end{equation}

\noindent Whereby, if only one meta-population is considered, the SIR model results are recovered \cite{nishiura2009effective}.

\subsection*{SEIIR-type meta-population model}

Finally, we present a meta-population model related to the transmission dynamics of SARS-Cov-2 coronavirus. As in the previous section, the meta-populations are connected by the commuter movement of individuals. In this model, individuals that are infected have to pass through a latency period to become infectious, during that time, we consider that those are in the exposed compartment $E$. After a mean latency period of $1/k$, those infected can become symptomatic or asymptomatic, $I^s$ and $I^a$ respectively. While in these compartments, the individuals of ``$j$'' are able to generate new infected ones in ``$i$'' based on the transmission rate $\lambda_{ij}$. However, asymptomatic ones are considered to have a lower transmissibility, thus, the transmission rate must be multiplied by a factor $\delta$ that lowers it's infectivity. The SEIIR model was developed by Oliveira in \cite{Oliveira2020mathematical} and in Supplementary Material 2 we derive an SEIIR-type meta-population model that can be written as:

\begin{align}
\frac{dS_i}{dt} =& -\sum_{j}^{n} \lambda_{ij}(t)\; S_i(t) \Big [I^s_j(t)+\delta I^a_j(t)\Big ] \; S_i(t),\label{eqS-met2}\\
\frac{dE_i}{dt}=& \sum_{j}^{n} \lambda_{ij}(t)\; S_i(t) \Big [I^s_j(t)+\delta I^a_j(t)\Big ] -\kappa \; E_i(t),\\
\frac{dI^s_i}{dt}=& p\kappa \; E_i(t) -\gamma_s I^s _i(t),\\
\frac{dI^a_i}{dt}=& (1-p)\kappa \; E_i(t) -\gamma_a I^a _i(t),\\
\frac{dR_i}{dt}=& \gamma_s I^s_i(t) + \gamma_a I^a_i(t).\label{eqR2}
\end{align}

To proceed with the methodology, we sort the compartments as $X=[ E_1, ..., E_n, {I_1}^{a}, ..., {I_n} ^{a}, {I_1} ^{s}, ..., {I_n} ^{s}]$ and $x=[ {i^e} _{1}, ..., {i^e} _{n}, {i^s} _{1}, ..., {i _n} ^{s}, {i_1} ^{a}, ..., {i_n} ^{a}]$ in a way that if there are $n$ meta-populations we have $3n$ infected compartments. Therefore, we obtain $\mathbfcal{F}(t)$ and $\mathbfcal{V}(t,\tau)$ as:

\begin{align}\label{F_SEIIR}
\mathcal{F}_i(t)= \begin{cases}
   \sum_{j}^{n} \lambda_{ij}(t)\; S_i(t) \Big [I^s_j(t)+\delta I^a_j(t)\Big ] , & \text{for $i\leq n$} \\
     0, & \text{for $i > n$}
    \end{cases}
\end{align}

\begin{align}\label{V_SEIIR}
\mathcal{V}_i(t,\tau)= \begin{cases}
    \kappa\; x_i(t,\tau) , & \text{for $i\leq n$} \\
    \gamma_s \; x_i(t,\tau) - p\kappa\; x_{i-n}(t,\tau), & \text{for $ n < i \leq 2n$} \\
    \gamma_a \; x_i(t,\tau) - (1-p)\kappa\; 
    x_{i-2n}(t,\tau), & \text{for $ 2n < i$}
    \end{cases}
\end{align}

\noindent where $I_j^s=X_{j+n}$ and $I_j^a=X_{j+2n}$. Thus, from \eqref{F_SEIIR}, we obtain:

\begin{align}\label{Omega_SEIIR}
\bm{\Omega}(t)= \begin{pmatrix}
    \bm{0}  & \rvline &\bm{\Omega^s} & \rvline & \bm{\Omega^a}   \\  \hline
        \bm{0}  & \rvline &\bm{0} & \rvline & \bm{0} \\ \hline  
        \bm{0}  & \rvline &\bm{0} & \rvline & \bm{0}
\end{pmatrix}, \qquad \qquad -\frac{\partial \bm{\mathbfcal{V}}}{\partial \bm{x}}= \begin{pmatrix}
     -\kappa & 0  & 0 \\
        p \kappa & -\gamma_s & 0 \\ 
          (1-p) \kappa & 0 & -\gamma_a
\end{pmatrix}  \mbox{\Large{$\otimes$}} \; \mbox{\Large{$\mathds{I}_n$}}.
\end{align}

\noindent Whereby ${\mathds{I}_n}$ is an  $n \times n$ identity matrix and $\otimes$ is a tensor product. Also, $\bm{\Omega}$ is divided into nine $n \times n$ submatrices, where $\bm{0}$ is an all-zeroes $n \times n$ matrix, $\bm{\Omega}^s\equiv \Big [\lambda_{ij} S_i \Big ]$ and $\bm{\Omega}^a\equiv  \delta  \bm{\Omega}^s$. Solving the system of differential equations on the characteristic line, we obtain: 

\begin{align}\label{Gamma_SEIIR}
\bm{\Gamma}(\tau)= \begin{pmatrix}
     e^{-\kappa \tau} & 0  & 0 \\[2ex]
        p\frac{\kappa}{\gamma_s -\kappa} \Big(e^{-\kappa\tau} - e^{-\gamma_s\tau} \Big ) & e^{-\gamma_s \tau} & 0 \\[2ex]
          (1-p)\frac{\kappa}{\gamma_a -\kappa} \Big(e^{-\kappa\tau} - e^{-\gamma_a\tau} \Big ) & 0 & e^{-\gamma_a \tau}
\end{pmatrix}  \mbox{\Large{$\otimes$}} \; \mbox{\Large{${\mathds{I}_n}$}}.
\end{align}

We proceed into the multiplication of the matrices $\bm{\Omega}$ and $\bm{\Gamma}$ in order to obtain:

\begin{align}\label{Omega_SEIIR}
\bm{A}(t,\tau)= \begin{pmatrix}
    \bm{A^e}  & \rvline &\bm{A^s} & \rvline & \bm{A^a} \\  \hline
        \bm{0}  & \rvline &\bm{0} & \rvline & \bm{0} \\ \hline  
        \bm{0}  & \rvline &\bm{0} & \rvline & \bm{0}
\end{pmatrix}
\end{align}

\noindent whereby the submatrices are defined as:

\begin{equation}
    \bm{A^e}(t) \equiv \Bigg [ \lambda_{ij}(t) S_i(t) \bigg(   p\frac{\kappa}{\gamma_s -\kappa} \Big(e^{-\kappa\tau} - e^{-\gamma_s\tau} \Big ) + \delta  (1-p)\frac{\kappa}{\gamma_a -\kappa} \Big(e^{-\kappa\tau} - e^{-\gamma_a\tau} \Big ) \bigg)   \Bigg ],
\end{equation}

\noindent $\bm{A^s}\equiv e^{-\gamma_s \tau}\bm{\Omega^s}$ and $\bm{A^a}\equiv e^{-\gamma_a \tau} \bm{\Omega^a}$. Therefore, the next generation matrix is

\begin{align}\label{Omega_SEIIR}
\bm{\mathbfcal{R}}(t)= \begin{pmatrix}
    \bm{\mathbfcal{R}^e}  & \rvline &\bm{\mathbfcal{R}^s} & \rvline & \bm{\mathcal{R}^a} \\  \hline
        \bm{0}  & \rvline &\bm{0} & \rvline & \bm{0} \\ \hline  
        \bm{0}  & \rvline &\bm{0} & \rvline & \bm{0}
\end{pmatrix}
\end{align}

\noindent where:

\begin{equation}
    \bm{\mathbfcal{R}^e}(t) \equiv \Bigg [ \lambda_{ij} (t) S_i(t) \bigg(   \frac{p}{\gamma_s}  + \delta  \frac{(1-p)}{\gamma_a} \bigg)   \Bigg ],
\end{equation}

\noindent Because there is no generation of infected individuals on the symptomatic and asymptomatic compartments, the expressions for $\bm{\mathbfcal{R}^s}$ and $\bm{\mathbfcal{R}^a}$ are not needed. This is so because $\mathcal{F}_i(t)$ for $i>n$ is null for any time. Therefore, the renewal equations of $\mathcal{F}_i(t)$ for $i>n$ have the tautological result that zero equals zero, regardless of the values of $\bm{\mathbfcal{R}^s}$ or $\bm{\mathbfcal{R}^a}$. Thus, all we need to describe the dynamics is $\bm{\mathbfcal{R}}$ for $i,j\leq n$, that is $\bm{\mathbfcal{R}^e}$. Lastly, we obtain the generation interval distribution matrix for $i,j\leq n$:

\begin{equation}
    g_{ij}(\tau) \equiv g(\tau)= \frac{\frac{p}{\gamma_s} g^s(\tau) + \frac{\delta (1-p)}{\gamma_a}g^a(\tau)}{\frac{p}{\gamma_s} + \frac{\delta (1-p)}{\gamma_a}},
\end{equation}

\noindent for

\begin{equation}
      g^a(\tau) = \frac{\kappa \gamma_a}{\gamma_a - \kappa} (e^{-\kappa \tau} - e^{-\gamma_a \tau} ), \qquad \qquad \qquad g^s(\tau) =  \frac{\kappa \gamma_s}{\gamma_s - \kappa} (e^{-\kappa \tau} - e^{-\gamma_s \tau} ) .
\end{equation}

\noindent Whereby, if only one meta-population is considered, we return to the results in literature \cite{Oliveira2020mathematical}.

\newpage
\topskip0pt
\vspace*{\fill}
\begin{center}
\begin{minipage}{.6\textwidth}
\Large{\textbf{Supplementary Material 2}}\\
\large{\textcolor{gray}{\textbf{Meta-population models formulation}}}
\end{minipage}
\end{center}
\vspace*{\fill}

\newpage

\section*{Meta-population models formulation}

In this Supplementary Material we will be, inspired by \cite{miranda2021scaling}, developing a meta-population model that takes into account the movement of individuals between meta-populations to describe the propagation of a disease through out multiple municipalities . We will be interpreting the inter-municipal flow as a complex network where its nodes represent municipalities and the weight of its edges represents the intensity of the flow between the connected municipalities. We also consider that each municipality can be interpreted as a meta-population, with its own compartments and parameters, which can be represented by a vector $\bm{y}_i(t)$, where the index “$i$” indicates the meta-population portrayed.Each entry of $\bm{y}_i(t)$ represents a compartment of this meta-population, such as $\bm{y}_i(t)=\big (S_i(t), I_i(t), R_i(t) \big)$ in the case of a SIR model. In which $S_i(t)$, $I_i(t)$ and $R_i(t)$ correspond to the susceptible, infected and removed individuals that are residents of the meta population "$i$", respectively. The sum of the elements of
$\bm{y}_i$ corresponds to the number of individuals of the meta-population "$i$", in the SIR model we have $S_i(t) + I_i(t) + R_i(t) =N_i(t)$.

We can represent the amount of individuals that goes from the meta-population "$i$" to another meta-population "$j$" each day as $\varphi_{ij}$, so it represents the flow of individuals between those meta-populations. Since each meta-population is described from it's compartments, we can describe the flow of those using $\bm{y}_i(t)$ in

\begin{equation}
    \text{Flow from $i$ to $j$} =  \varphi_{ij}(t) \frac{\bm{y}_{i}}{N_i}.\centering
\end{equation}

We define $\Phi_{ij}(t) \equiv \frac{\varphi_{ij}(t)}{N_i}$ as the density of flow. In that way, $\Phi_{ij}\; \bm{y_i}$ is the number of individuals from each compartment class that are flowing from "$i$" to "$j$". It's natural to see that sum of $\Phi_{ij}\; \bm{y}_i$ for each compartment class of $\bm{y}_i$ is equal to $\varphi_{ij}$, in the SIR model for example: $\Phi_{ij}\; S_i +\Phi_{ij}\;I_i + \Phi_{ij}\; R_i = \varphi_{ij}$.

Given that the meta-populations are connected through the flow , it is necessary to identify how they. For this purpose, graphs are used which, in general, represent complex networks, given the large number of connections. To represent networks, matrices are used, as in the case of adjacency matrices whose elements represent the density of flow ($\Phi_{ij}$). Thus, the values of each $\Phi_{ij}(t)$ are the weight of the edges with a null value representing the absence of an edge. Since there are no self interactions in this network, we do not consider the flow of a meta-population to itself, thus  $\varphi_{ii}\equiv 0$ .

Due to the circulation of individuals through the network, the characteristics of the meta-populations will be changed. Thus, we define the effective population of ``$i$'' described by the vector $\bm{y}_{e,\;i} = \big(S_{e\; ,i}(t),I_{e\; ,i}(t),R_{e\; ,i}(t) \big)$ which corresponds to individuals located, at time t, in the “i” meta-population regardless of their origin meta-population.The effective population represents the new characteristics of a population due to the flow; for example, a meta-population that does not have infected resident individuals may have carriers of the pathogen in their effective population due to the flow from some other meta-population that presents infected individuals. We can write $\bm{y}_{e,\; i}$ as:

\begin{equation}
    \bm{y}_{e,\; i}(t) = \overbrace{\bm{y}_i(t)}^\text{ Resident Pop. } -\quad\overbrace{ \sum_{j \neq i}^{n} \, \Phi_{ij} \; \bm{y}_{i}(t) }^\text{Outflow} \quad+ \quad  \overbrace{\sum_{j \neq i}^{n} \Phi_{ji} \;\bm{y}_{j}(t)}^\text{Inflow}
\end{equation}

For, $N$ being the number of meta populations. Rearranging the equation:

\begin{equation}\label{Effective_Pop}
     \bm{y}_{e,\;i}(t) = \bm{y}_{i}(t)\Big( 1 - \sum_{j}^{n} \, \Phi_{ij} \Big) + \sum_{j}^{n} \Phi_{ji}\; \bm{y}_{j}(t)
\end{equation}

It's clear then that the effective population of "$i$" is the sum of the individuals from "$i$" that are in "$i$" , $\bm{y}_{i}\Big( 1 - \sum_{j \neq i}^{n} \, \Phi_{ij} \Big)$, with the individuals from other meta populations that are in "$i$",  $\sum_{j \neq i}^{n} \Phi_{ji}\; \bm{y}_{j}$. If we sum each element of $\bm{y}_{e,\; i}$ we have the effective population number: $N_{e,\; i}= N_i - \sum_{j \neq i}^{n}\varphi_{ij} + \sum_{j \neq i}^{n}\varphi_{ji}$. It is important to highlight that the resident population is not changed, therefore, we do not consider migration but only commuter periodic movement, whereby the individuals of one meta-population go to another to work or study.

We now proceed in order to describe how the transmission of a disease occurs in a meta-population, taking into account the flow of individuals in the network. We must then describe the occurrence of new infections in the meta-population $\bm{y}_i(t)$, which can occur within the “i” meta-population

\begin{equation}\label{Infection_In_city}
    \frac{\beta _i}{N_i}  \, \times 
\overbrace{S_{i} \Big( 1 - \sum_{j}^{n} \, \Phi_{ij} \Big)}^\text{Susceptible Indv. from ``$i$'' in ``$i$''}  \times \overbrace{I_{e,\;i}}^\text{Infected Indv. in ``$i$''},
\end{equation} 

\noindent or on another meta-population "$j$"

\begin{equation}\label{Infection_Out_city}
    \frac{\beta _j}{N_j} \times \overbrace{\Phi_{ij} \,  S_{i}}^\text{Susceptible Indv. from ``$i$'' in ``$j$''}  \times \overbrace{I_{e,\; j}}^\text{Infected Indv. in ``$j$''}.
\end{equation}

\noindent $\beta_i(t)$ and $\beta_j(t)$ being the transmission rates within the “i” and “j” meta-populations, respectively. The parameters $\beta_i(t)$ and $\Phi_{ij}(t)$ are time dependent, so we can incorporate the changes in the behavior of the populations on those variables. Thus, the number of new infections of individuals that belong to “$i$”, $\mathcal{F}_i(t)$, will be equal to the sum of \eqref{Infection_In_city} and \eqref{Infection_Out_city}. We can substitute equation \eqref{Effective_Pop} in expressions \eqref{Infection_In_city} and \eqref{Infection_Out_city}, in order to expand the term of the effective population of infected individuals($I_{e\; ,i}(t)$ ,$I_{e\; ,i}(t)$) and be able to express them according to the infected residents in each meta-population. Carrying out these operations, we get to:

\begin{equation}\label{Lambda}
\mathcal{F}_i(t) =  \sum_{j}^{n} \lambda_{ij}(t)\; S_i(t)I_j(t)
\end{equation}

For:
\begin{align}
    \lambda_{ii}&=\frac{\beta _i}{N_i} \; \Big( 1 - \sum_{j}^{n} \, \Phi_{ij} \Big)^2 + \sum_{j}^{n}\;  \frac{\beta _j}{N_j} \; \Phi_{ij}^2 \label{lambdaI} \\
    \lambda_{ij}&= \frac{\beta _i}{N_i} \; \Phi_{ji} \Big( 1 - \sum_{k}^{n} \, \Phi_{ik} \Big) + \frac{\beta _j}{N_j}  \; \Phi_{ij} \; \Big( 1 - \sum_{k}^{n} \, \Phi_{jk} \Big) + \sum_{k}^{n} \frac{\beta _k}{N_k} \Phi_{ik} \; \Phi_{jk} \label{lambdaJ}
\end{align}

The $\lambda_{ij}(t)$ are related to the transmission between individuals of the same meta-population and from individuals of the “j” meta-population to individuals of the “i” meta-population.

\subsection*{SIR-Type model}

In this section, we construct the set of equations for the SIR-type model based in the contamination process modeled in this Supplementary Material. We will be modeling the resident populations of each municipality and will not be considering migration, therefore, resident individuals of one meta-population will always remain residents of that same meta-population. It is assumed that the susceptible population can only decrease due to infection process, therefore, no life and death dynamic. In the same way, the removed individuals are only generated by a removing rate $\gamma$, uniform for all meta-populations. Gathering those considerations, we write the system of equations:

\begin{align}
\frac{dS_i}{dt} =& - \sum_{j}^{n} \lambda_{ij}(t)\;I_j(t) \; S_i(t),\label{eqS-met}\\
\frac{dI_i}{dt}=& \sum_{j}^{n} \lambda_{ij}(t)\;I_j(t)\; S_i(t) -\gamma \; I_i(t),\label{eqI}\\
\frac{dR_i}{dt}=& \gamma I_i(t).\label{eqR}
\end{align}

\noindent Where it is clear that when $\varphi_{ij}=0$, for all $i$'s and $j$'s, each meta-population will be described by the classical SIR homogeneous model. Of course, this SIR approach is not taking into account the other heterogeneities of a disease besides space. On the following section we present a more sophisticated approach focused on the Covid-19 transmission dynamics. 

\newpage

\subsection*{A meta-population model for Covid-19 (SEIIR)}
In this section, we establish a more precise description for the dynamics of SARS-Cov-2 coronavirus on the municipalities network. To do so, we consider, as in \cite{Oliveira2020mathematical}, that the infected individuals can be separated in three classes: the exposed $E$, individuals
infected which are in the latency period and do not transmit the disease; the symptomatic individuals $I^s$, that are infectious, present a substantial amount of symptoms and are registered in the official data; the 
asymptomatic/undetected ones $I^a$, that are infectious but present mild/non-existing symptoms and are not registered in the official data. The infected individuals always start in the exposed compartment, a portion $p$ of them eventually becomes symptomatic and the other $(1-p)$ portion becomes asymptomatic. Since we have two types of infectious individuals, both of them must be taken in consideration in the generation of new infected. Therefore, assuming that the asymptomatic individuals transmission rate is a fraction $\delta$ of the symptomatic transmission rate, it is simple to derive, similarly to \eqref{Infection_In_city} and \eqref{Infection_Out_city}, the number of infected individuals that are generated on the exposed compartment of each meta-population:

\begin{equation}\label{Lambda2}
\mathcal{F}^e_i(t) =  \sum_{j}^{n} \lambda_{ij}(t)\; S_i(t) \Big [I^s_j(t)+\delta I^a_j(t)\Big ].
\end{equation}

\noindent Therefore, by the same assumptions presented on the previews section, the following model is obtained:

\begin{align}
\frac{dS_i}{dt} =& -\sum_{j}^{n} \lambda_{ij}(t)\; S_i(t) \Big [I^s_j(t)+\delta I^a_j(t)\Big ] \; S_i(t),\label{eqS-met2}\\
\frac{dE_i}{dt}=& \sum_{j}^{n} \lambda_{ij}(t)\; S_i(t) \Big [I^s_j(t)+\delta I^a_j(t)\Big ] -\kappa \; E_i(t),\\
\frac{dI^s_i}{dt}=& p\kappa \; E_i(t) -\gamma_s I^s _i(t),\\
\frac{dI^a_i}{dt}=& (1-p)\kappa \; E_i(t) -\gamma_a I^a _i(t),\\
\frac{dR_i}{dt}=& \gamma_s I^s_i(t) + \gamma_a I^a_i(t).\label{eqR2}
\end{align}

\noindent Whereby $\kappa$, $\gamma_s$ and $\gamma_a$ are the removing rate of the exposed, symptomatic and asymptomatic compartments, respectively, and are uniform in all meta-populations.

\newpage
\topskip0pt
\vspace*{\fill}
\begin{center}
\begin{minipage}{.6\textwidth}
\Large{\textbf{Supplementary Material 3}}\\
\large{\textcolor{gray}{\textbf{Expressions and parameter values for evaluating $\mathbfcal{R}(t)$ for the meta-population models}}}
\end{minipage}
\end{center}
\vspace*{\fill}

\newpage

\section*{Expressions for evaluating $\mathbfcal{R}(t)$ using incidence data}

Here, we derive the expressions and parameter values needed to estimate the reproduction numbers for the meta-population models. Firstly, we start by substituting the explicit expressions of the reproduction numbers of each model in the equation 2.24 of the main framework. After some straightforward calculations, we obtain:

\begin{equation}\label{Est}
    \mathbfcal{Q}_i (t) = \sum_j ^n \lambda_{ij}(t) a_j(t).  
\end{equation}

\noindent whereby, $\mathbfcal{Q}_i (t) = \mathcal{B}_i(t)/S_i(t)$ and 

\begin{equation}
    a_j= \sum_{\tau=0} ^t \frac{1}{\gamma} g(\tau) \Delta t
\end{equation}

\noindent for the SIR model and 

\begin{equation}
    a_j= \sum_{\tau=0} ^t \Big[ \frac{p}{\gamma_s} + \frac{\delta (1-p)}{\gamma_a}\Big] g(\tau) \Delta t
\end{equation}

\noindent for the SEIIR model. We consider that the $\mathcal{B}_i(t)$ is the collection of all the new infections during a $\Delta t$ time interval of 1 day. With appropriate units of measure, we have $\Delta t=1$. In the SIR model we consider that every infection is reported, $\rho_i=1$ and in the SEIIR case only the symptomatic individuals report their infections, $\rho_i=p$. We proceed into estimating the susceptible population. If we integrate \eqref{eqS-met} and substitute \eqref{Lambda} and equation 2.23 of the main framework, we get:

\begin{equation}
    S_i(t)= N_i - \sum_{t'=0} ^t\frac{\mathcal{B}_i(t')}{\rho_i}.
\end{equation}

 Finally, substituting \eqref{lambdaI} and \eqref{lambdaJ} in \eqref{Est} results in:

\begin{equation}\label{Est2}
    \mathbfcal{Q}_i (t) = \sum_j ^n \Theta_{ij}(t) \beta_j(t), 
\end{equation}

\noindent for:

\begin{align}
    \Theta_{ii}&= \frac{1}{N_i} \Bigg[ a_i \Big( 1- \sum_k ^n \varphi_{ik} \Big )^2 + \sum_j ^n a_j \varphi_{ji} \Big(1-\sum_k ^n \varphi_{ik} \Big ) \Bigg ],\label{ThetaI} \\
    \Theta_{ij}&= \frac{1}{N_j} \Bigg[ a_j \varphi_{ij} \Big(1-\sum_k ^n \varphi_{jk} \Big ) + \varphi_{ij} \sum_{k} ^{n} a_k \varphi_{kj}\Bigg ].
    \label{ThetaJ}
\end{align}

\noindent Therefore value of $\mathcal{Q}_i(t)$ and $\theta_{ij}$ for each day can be obtained using the reported data and parameters. Thus, for each day, this leaves us with algebraic system of ``$n$'' variables, $\beta_j(t)$ and ``$n$'' equations. We can analytically solve this system for every meta-population, with the help of a computer algorithm, and obtain the daily values of $\beta_j(t)$ of every meta population. Substituting the values of every $\beta_j(t)$ in \eqref{lambdaI} and \eqref{lambdaJ}, we can compute the values of the $\lambda_ij(t)$'s which leads us to the values of the reproduction number, equations 3.1 and 3.2 of the main framework. Since the available data is of daily number of cases, the $\bm{\mathcal{R}}(t)$ is evaluated for each calendar day. 

\subsection*{Parameters}

To estimate the $\beta_i(t)$ parameters, and consequently estimate the reproduction numbers, the parameters of both models must be obtained. $\delta$, $\gamma_a$, $\gamma_s$, $p$, and $\kappa$ are estimated for the state of Rio de Janeiro in \cite{jorge2020assessing} and can be found on Table \ref{tabPara}. Additionally, in the SIR type model we assume $\gamma=\gamma_s$. The intermunicipal commuter movement of workers and students for the cities of Brazil, $\varphi_{ij}$, can be found in a study conducted by IBGE (Brasilian Institute of Geography and Statistics) \cite{ibge2016arranjos}.

\begin{table}[H]
\caption{Key epidemiological parameters of the SEIIR model obtained in \cite{jorge2020assessing}. }
\label{tabPara}
\centering
\begin{tabular}{lllllll}
\hline
\textbf{Parameter} & \textbf{Description} & \textbf{Value}  
\\ \hline
$\delta$    & Asymptomatic/non-detected infectivity factor  & $0.258$   
\\
$p$     & Proportion of latent (E) that proceed to symptomatic infective & $0.273$
\\
$\kappa^{-1}$    & Mean exposed period (days$^{-1}$) & $1/0.25$    
\\
$\gamma_{a}^{-1}$   & Mean asymptomatic period (days$^{-1}$)  & $1/0.288$  
\\
$\gamma_{s}^{-1}$  & Mean symptomatic period (days$^{-1}$) & $1/0.25$  

\\ \hline
\end{tabular}
\end{table}

\newpage
\topskip0pt
\vspace*{\fill}
\begin{center}
\begin{minipage}{.6\textwidth}
\Large{\textbf{Supplementary Material 4}}\\
\large{\textcolor{gray}{\textbf{Additional information about the municipalities}}}
\end{minipage}
\end{center}
\vspace*{\fill}

\newpage

\section*{Additional information about the municipalities}

\begin{table}[H]
\centering
\resizebox{0.9\textwidth}{!}{%
\begin{tabular}{llll}
\hline
\textbf{Municipality}   & \textbf{Acronyms} & \textbf{Populational size} & \textbf{Total reported cases} \\ \hline
Belford Roxo            & BR                & 485,687                    & 8,578                         \\
Duque de Caxias         & DdC               & 905,129                    & 8,736                         \\
Magé                    & Ma                & 242,113                    & 3,539                         \\
Mesquita                & Mq                & 800,835                    & 1,351                         \\
Nilópolis               & Ns                & 154,749                    & 1,245                         \\
Niterói                 & Nt                & 497,883                    & 12,165                        \\
Nova Iguaçu             & NI                & 167,287                    & 5,908                         \\
Queimados               & Q                 & 150,333                    & 2,376                         \\
\textbf{Rio de Janeiro} & RJ                & 6,592,227                  & 95,444                        \\
São Gonçalo             & SG                & 1,075,372                  & 11,601                        \\
São João de Meriti      & SJdM              & 448,340                    & 3,167                         \\ \hline
\end{tabular}%
}
\caption{}
\caption{ The selected cities of the state of Rio de Janeiro in the Southeast of Brazil with their acronyms, number of inhabitants, and the reported COVID-19 cases until September 14th (2020). In bold, the capital (Rio de Janeiro).}
\label{tab:my-table}
\end{table}

\begin{figure}[h]
     \centering

    \includegraphics[width={\linewidth}]{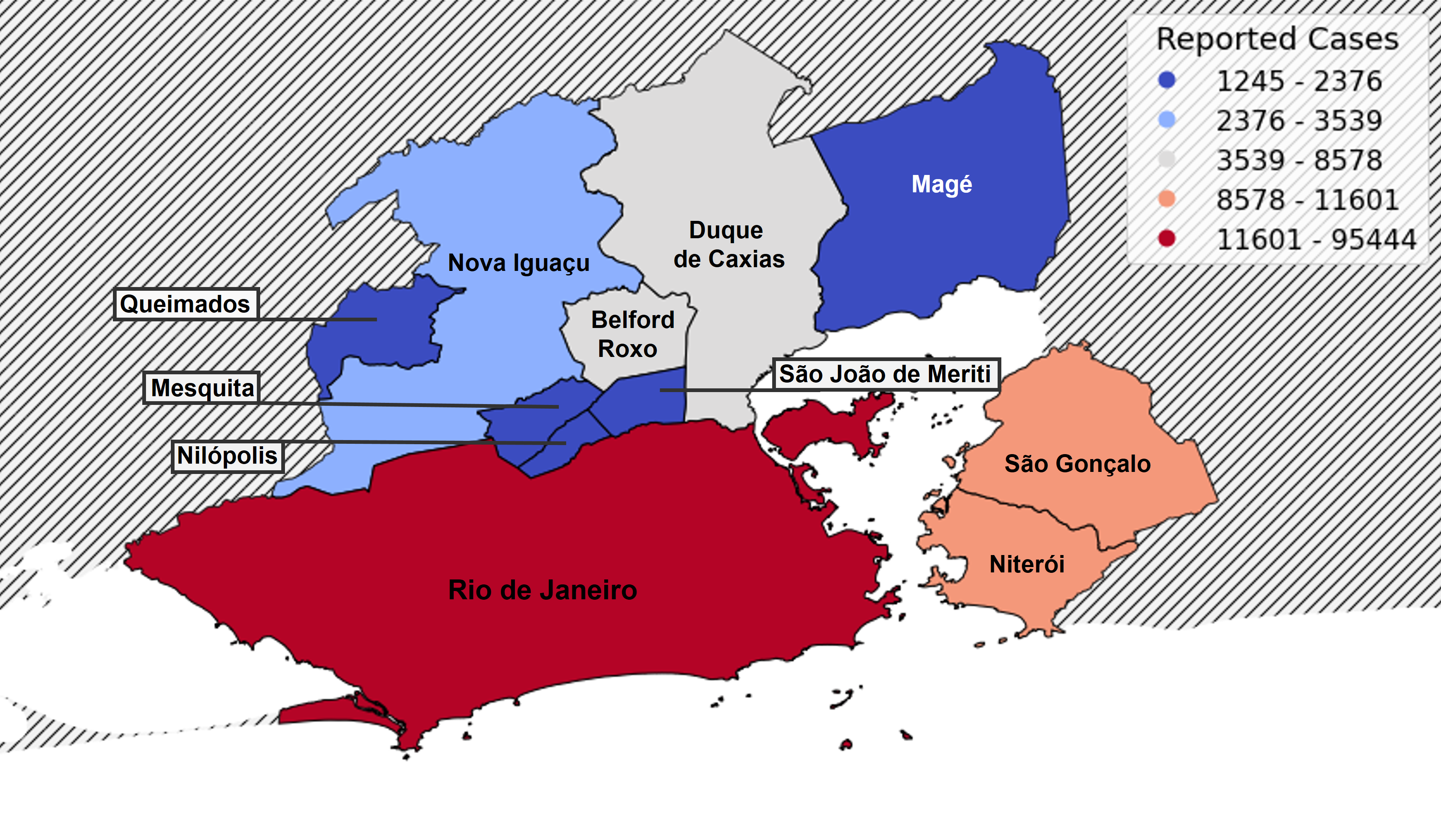}

    \caption{Distribution of cases by the chosen cities of this study.}
        \label{map}
\end{figure}

\begin{figure}[H]
     \centering

    \includegraphics[width={\linewidth}]{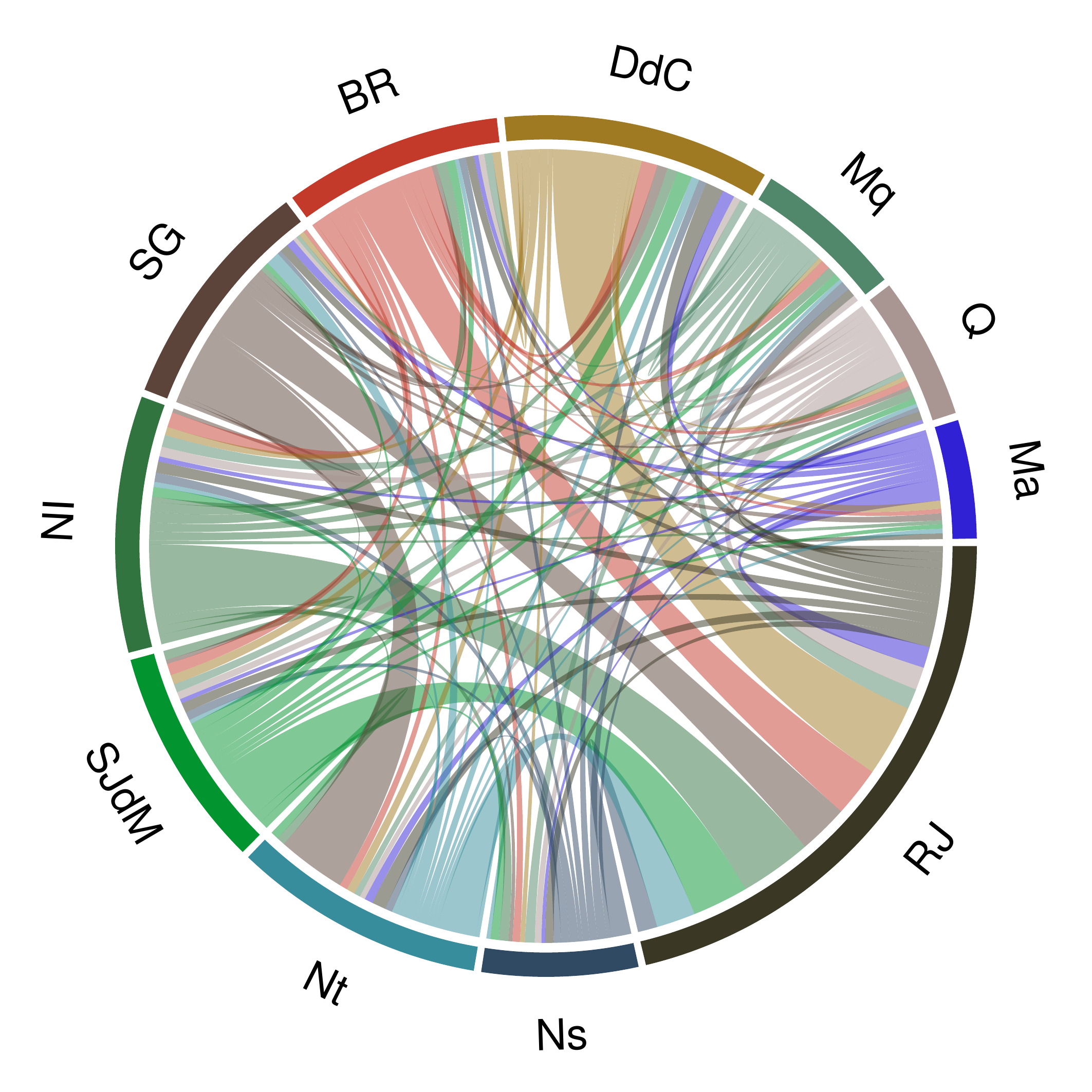}

    \caption{Average daily commuter movement between municipalities, due to workplaces. The thickness of each line is proportional to the amount of movement from one municipality to another. Colors are used only for the purpose of better visualization and have no specific meaning.}
        \label{map}
\end{figure}

\newpage
\bibliography{sample}